\documentclass[10pt,journal,compsoc]{IEEEtran}

\IEEEoverridecommandlockouts
\usepackage{graphicx}
\usepackage{amsmath}
\usepackage{amsthm}
\usepackage{subfigure}
\usepackage{epsfig}
\usepackage[ruled,linesnumbered]{algorithm2e}
\usepackage{color, soul}
\usepackage[table,xcdraw]{xcolor}
\graphicspath{{eps/}}
\newtheorem{theorem}{Theorem}

\newcommand{\rthree}[1]{\definecolor{yellow}{RGB}{255,255,0}\sethlcolor{yellow}\hl{\textbf{#1}}}

\ifCLASSINFOpdf

\else

\fi

\begin{document}

\title{RRFT: A Rank-Based Resource Aware Fault Tolerant Strategy for Cloud Platforms}

\author{Chinmaya~Kumar~Dehury,~\IEEEmembership{Member,~IEEE}, Prasan~Kumar~Sahoo,~\IEEEmembership{Senior Member,~IEEE} and Bharadwaj~Veeravalli,~\IEEEmembership{Senior Member,~IEEE}  

\thanks{Chinmaya Kumar Dehury was with the department of Computer Science and Information Engineering, Chang Gung University, Guishan, Taiwan.
	He is currently with Mobile \& Cloud Lab, Institute of Computer Science, University of Tartu, Estonia. 
	Email: chinmaya.dehury@ut.ee}

\thanks{Prasan Kumar Sahoo (Corresponding Author) is with department of Computer Science and Information Engineering, Chang Gung University, Guishan 333, Taiwan. He is an Adjunct Research Fellow in the Department of Neurology, Chang Gung Memorial Hospital, Linkou 33305, Taiwan. Email: pksahoo@mail.cgu.edu.tw}

\thanks{Bharadwaj Veeravalli is with Department of Electrical and Computer
Engineering, National University of Singapore, Singapore. Email: elebv@nus.edu.sg}}

\markboth{IEEE Transactions on Cloud Computing,~Vol.~xx, No.~x, month~year}%
{Shell \MakeLowercase{\textit{et al.}}: Bare Demo of IEEEtran.cls for Computer Society Journals}

\IEEEtitleabstractindextext{%
\begin{abstract}
The applications that are deployed in the cloud to provide services to the users encompass a large number of interconnected dependent cloud components. Multiple identical components are scheduled to run concurrently in order to handle unexpected failures and provide uninterrupted service to the end user, which introduces resource overhead problem for the cloud service provider. Furthermore such resource-intensive fault tolerant strategies bring extra monetary overhead to the cloud service provider and eventually to the cloud users. In order to address these issues, a novel fault tolerant strategy based on the significance level of each component is developed. 
The communication topology among the application components, their historical performance, failure rate, failure impact on other components, dependencies among them, \rthree{etc.}, are used to rank those application components to further decide on the importance of one component over others.
Based on the rank, a Markov Decision Process (MDP) model is presented to determine the number of replicas that varies from one component to another.
A rigorous performance evaluation is carried out using some of the most common practically useful metrics such as, recovery time upon a fault, average number of components needed, number of parallel components successfully executed, etc., to quote a few, with similar component ranking and fault tolerant strategies. Simulation results demonstrate that the proposed algorithm reduces the required number of virtual and physical machines by approximately 10\% and 4.2\%, respectively, compared to other similar algorithms.
\end{abstract}
}

%
\maketitle
\IEEEdisplaynontitleabstractindextext
\IEEEpeerreviewmaketitle

\section{Introduction}
Huge amount of physical servers connected through very high network bandwidth is the backbone of cloud computing environment. The pay-as-you-go model of cloud computing provides the flexibility for users to access physical servers on rented basis from cloud service provider (CSP). The resources are provided through different service models such Software as a Service (SaaS), Platform as a Service (PaaS), and Infrastructure as a Service (IaaS) \cite{Ghazouani201761, dehurydyvine}. For example, Microsoft Office 365 is implemented in SaaS model to provide the MS office software through the web.

The applications that are specifically designed to deploy in the
cloud are referred to as cloud applications. Such applications are
accessed remotely through multiple web pages \cite{Ding201747}. Network bandwidth plays an
important role in providing the software functionalities to the
users. In order to simplify the cloud application at the
development stage, the entire cloud application is divided into
multiple small applications, known as cloud component. A cloud
application component is a part of the software package or a
module that encapsulates a set of interrelated functions or tasks.
For example, in document editing cloud applications, the module
that is responsible for saving and exporting the document in a
specific format can be considered as one cloud component.
Each cloud component receives a set of inputs and executes a set
of predefined operations and provides the output to either the
user or to other dependent components. In the case of ATM services
provided by the banks to its customer, reading the ATM card,
communicating the desired remote database, printing the receipt,
etc., can be considered as different dependent components.


On the other hand, fault tolerance is becoming an important issue
for the cloud research community. The CSP may experience
unexpected failure at any time while providing the service through efficient scheduling mechanisms \cite{Firmament2016}. A fault may
occur at the hardware level due to a power outage or at the software
level \cite{FTworkflow_2017}, which may occur due to the ambiguous
input to the cloud application. Different mechanisms have been
proposed in recent years to prevent and handle the pre-failure and
post-failure situations. Authors in \cite{cheraghlou2016survey}
present an extensive survey on fault tolerant architectures in
cloud computing.

Different fault tolerant mechanisms that are available in the research
community are \textit{recovery block}, \textit{N-version
programming}, \textit{parallel}, and \textit{VM-restart}
\cite{7582315}. In order to prevent unexpected
failure, the applications are executed on multiple cloud servers.
In the recovery block fault tolerant mechanism, applications
are executed sequentially for a certain number of time until the
desired output is obtained. In contrast, the application is
scheduled to execute on multiple cloud servers concurrently in
parallel fault tolerant strategy until the first output is
obtained \cite{7742909}. Taking software bugs into
consideration, multiple equivalent applications are scheduled to
run concurrently on multiple cloud servers. All outputs are then
compared and the most similar output is provided to the user.
As the failure may occur at the virtual machine or cloud servers
level, restarting the virtual machine itself is one of the most
popular and simple fault tolerant strategies. In fault tolerance
strategies, a large number of resources are wasted, where multiple
VMs are engaged for a single component, especially for
long-running jobs \cite{7096978}. Among all fault tolerant
strategies, the time required to finish execution of the cloud
application and the amount of resource required are inversely
proportional to each other. In recovery block fault tolerant
strategy, the time required to finish the execution of application
successfully is more than that of the parallel fault tolerant
strategy and vice-versa.

Efficient fault tolerant strategies are mainly either time-driven or resource-driven. In case of time-driven strategies, the applications are executed multiple times until the required result is obtained. However, in the case of resource-driven, multiple identical applications or similar applications are executed by allocating more resources. Based on the critical level of the application, the CSP needs to provide sufficient amount of resource or time. If the application is time-constrained, CSP needs to provide more resource for parallel execution of the identical applications and ensure that the application is finished on time. However, if the application is cost-constrained, the CSP provide minimum but sufficient amount of memory to the application to execute and obtain the result. In such cases, the CSP executes the applications only upon failure. However, in the former case, irrespective of the correctness of the result, the CSP needs to allocate more than the required amount of memory. 

In the current research era, a huge number of complex,
time-consuming algorithms are designed to carry out specific
operations such as DNA sequencing in bioinformatics involve very
complex algorithms. As the current research is heading toward
simplifying the human life, the cloud applications are becoming
more complex and large in terms of the number of cloud components
involved in the corresponding applications. Components that are
involved in one cloud application are interconnected and dependent
on each other. Despite numerous advantages
provided by the cloud platforms, CSPs experience unexpected
failures in providing the services, which are the major research
issues. By connecting the unexpected failure issues to a large
number of complex inter-dependent cloud components, it is observed
that a single failure of any cloud component may lead to the
failure of the entire cloud application \cite{zheng2012component}. Failure of cloud applications
has a great impact on QoS which affects the decision to take the
cloud service from a particular CSP \cite{Ding201747}. On the
other hand, providing service with a higher degree of fault tolerance
is a resource-intensive job for cloud service providers, which
further introduces extra monetary overhead to the CSP and
eventually to the cloud users.

\subsection{Motivation} \label{sec:motivation}
The large cloud applications are mainly consisting of a large number of interconnected components, where each component is responsible for providing separate functionality. However, the user may experience interruption due to unexpected error or fault in the cloud application during service period, which incurs huge monetary cost to the cloud user as well as to the cloud service provider.

Cloud application may encounter faults due to several reasons such
as software bugs, erroneous input data, hardware failures, etc. 
A fault may bring down the whole cloud application or multiple components. 
For example, failure of the components that pre-process the incoming raw CCTV footage could hamper the whole real-time security system in a cloud-based smart city solution. On the other hand, some components are responsible for sending the processed and compressed CCTV footage to the storage servers for the backup purpose. Failure of such components would not affect the real-time security system. This paper mainly considers such cloud applications.
In general, at any particular time, a subset of
components is active providing services to
the users. It is assumed that components are associated with
different priorities based on the functionalities. Applying fault
tolerant mechanisms to the whole cloud application or all
components without considering the functionality of each component
could be resource-intensive job and therefore would be expensive.

The faults are handled by mainly two kinds of strategies: replication and re-execution. In the replication strategy, multiple identical components are scheduled to run concurrently, which is a resource-intensive strategy. In the re-execution strategy, the same component is executed again only if the component failed while providing the services. This strategy eliminates the additional resource demand. However, this introduces additional time as the component may need to start from either the beginning or from the last saved restore point.

Besides, the problem of providing a huge amount of resources to the
cloud application, the placement of the backup component plays a
vital role in providing an efficient fault tolerance strategy.
Primary components communicate among each other while providing
the service to the user. Furthermore, to provide uninterrupted
service by tolerating a higher degree of faults, primary components
communicate with their backup components in case of a replication strategy. Upon failure of the primary component, one of the backup components will act as the new primary
component and interact with other primary components. 
The network latency between the primary and the backup component may affect the QoS, which infers the importance of placement of both components. 
Here, placement of the component refers to the selection of a
physical machine for deploying the component. It is
essential to consider the communication or the placement of the
backup components as a major factor while designing the fault
tolerant strategy. Further, it is also not recommended to apply the re-execution strategy onto all the components in order to save the resources and minimize the service cost, as this would incur higher degradation of service quality. The aforementioned scenarios of providing
efficient fault tolerant service by reducing the total amount of
required resource without compromising the service quality and placement of component onto suitable PM
motivates us to propose an efficient fault tolerant mechanism.

As it is clearly mentioned in Section \ref{sec:motivation}, providing fault tolerant service to cloud application that consists of a very large number of cloud components is resource-intensive, in this paper, we attempt to propose a novel resource-aware strategy that provides fault tolerant services to the significant selective components instead of the whole application. Thus we focus on the design, analysis and evaluation of certain important metrics and experimenting with any real-life application is beyond the scope of this work and serves as an immediate extension, as discussed in Section \ref{sec:conclusion}.

The rest of this paper is organized as follows. The related literature
survey is presented in Section \ref{sec:relatedwork}. The concerned
problem is formulated in Section \ref{sec:probFormulation}. The
proposed solution that includes the determination of most
significant component followed by the fault tolerant algorithm is
presented in Section \ref{sec:solution}. The proposed optimization
policy for the total numbers of replicas is presented in Section
\ref{sec:performance}. Performance evaluation and concluding
remarks are made in Section \ref{sec:Perf_Evaluation} and
\ref{sec:conclusion}, respectively.

\section{Related Works}
\label{sec:relatedwork} An intensive survey and comparison of the
different fault tolerant architectures are presented in
\cite{KUMARI2018, HASAN2018156, cheraghlou2016survey}. It is emphasized that
despite numerous advantage of cloud computing such as reduction of
costs, efficient resource utilization \cite{dehurylvrm}, leveraging the efficiency and compatibility of software,
increasing the storage capacity, etc., immature fault tolerant
mechanism plays an important role affecting the decision to adopt
cloud computing environment. The survey in
\cite{cheraghlou2016survey} classifies the fault tolerance
architecture into proactive and reactive architecture. The fault tolerance techniques are applied in different stages of cloud computing, such as in scheduling \cite{8036409, 2018-fault-scheduling, SETLUR202014, yao2017using}, resource allocation \cite{2018-fault-TPDS, dean2016perfcompass}, to improve the reliability \cite{8964469}, placement of virtual network function in cloud network \cite{YUAN2020106953} etc.  

In the scenario of mobile cloud computing, where the entire
computation or the part of the computation payload is offloaded to
powerful cloud servers, authors in \cite{deng2015computation}
address the offloading problem by taking the portability of mobile
device and connectivity issues of the mobile network into
consideration. The fault tolerant based offloading mechanism is
designed based on the genetic algorithm. Despite numerous
advantages of genetic algorithms, the time-consuming process may
provide the extra computation overhead to the mobile device, and
eventually this may lead to the failure of proposed offloading
mechanism. Similar to \cite{zheng2012component}, authors in
\cite{qiu2014reliability} apply a similar concept to provide fault
tolerant service to different component of a cloud application.
The rank of a cloud component depends on the reliability
properties such as failure rate and failure impact. The fault
tolerant strategies are applied to the components based on the
reliability, response time, and resource cost. However, it is
observed that the component with higher failure rate and failure
impact in some component DAG topology are given lower rank value,
which contradicts the proposed ranking concept.

Authors in \cite{2018-fault-scheduling} proposed the hybrid task
scheduling mechanism offering the fault tolerant service by
integrating the tradition backup and checkpoint technology and
classifying the tasks and VMs. However, the proposed scheduling
strategy allocates the resources among all tasks specifically for
the backup purpose without investigating the importance of the
tasks. As a result, the resource required is very high in order to
implement the proposed scheduling mechanism. Similarly, in
\cite{2018-fault-TPDS}, the proposed fault tolerant mechanism
allocates maximum amount of resource to the tasks, which is a
resource-intensive strategy for the cloud service provider.
Similarly, authors in \cite{2018-cost-fault} formulate the
cost-effective fault tolerant strategies in providing services to
the multiple tenants in cloud. The major pitfall of the proposed
fault tolerant strategy is that the historical behavior of each
service is not taken into account, which could have given insight
behavior of each service in terms of the failure probabilities.

In order to handle the post-fault events, it is very much
essential to detect the fault in every physical server with no
delay, as described in \cite{smara2016acceptance}. Here, the authors
proposed a recovery block's acceptance test based fault detection
scheme for component-based cloud computing environment. The
proposed fault detection scheme is dedicated to only software
faults, transient hardware faults, and response-time failures.
However, the proposed idea does not address the issue of power
failure that may cause the failures of the entire server.

Though cloud provider guarantees the resource demand, however, in
the data center performance anomalies or failure of expected
performance is an upcoming issue, which arises due to sharing of
physical resources and multi-tenancy as discussed in
\cite{dean2016perfcompass}. To debug the application failure, distinguishing the faults with
global and local impact is essential, as in
\cite{dean2016perfcompass}.
Such classifications of faults provide
power to the system administrators and the application developer
to understand and diagnose the root cause of the failure. Authors
in \cite{wang2016fd4c}, extend the fault classification work as in
\cite{dean2016perfcompass} for automatic fault diagnosis of web
applications in cloud computing. Fluctuating workload, management
of large-scale application, and modeling the behavior of complex
application are some of the major factors that possess greater
challenges in fault diagnosis of web applications. The behavior of
access patterns is analyzed using correlation analysis, which is
used further to detect the faults. However, finding the
correlation between the workload and the application may lead to a
time-consuming process as the frequency arrival of the application
to the cloud service provider is very high.

Authors in \cite{chen2015energy} proposed fault tolerant enabled
storage and processing of data in mobile cloud taking energy
consumption as a major parameter into consideration. As running an
application needs processing and storage capabilities, authors
address the issue of selecting a suitable node that is a mobile
device or a cloud, for processing and data storage by following
\textit{k-out-of-n} computing approach. Here, $k$ and $n$
determine the degree of reliability. However, the proposed scheme
does not address the issues generated due to the mobility of peer
mobile nodes.


Authors in \cite{zheng2012component} propose the ranking of the
cloud components in order to find out the significant component. A
cloud application comprises a set of components represented as a
directed acyclic graph. The ranking of the cloud component depends
on the number of components present in the cloud application and
the number of components that invoke the corresponding component.
As a cloud application is responsible for providing certain services that are
consisting of  a large number of components, it is a hectic job to
provide the critical status to each component. Further, it is not
clear, which component should be given what kind of fault tolerant
service.

The major upcoming issue in resource allocation is to distribute
the resources among different users with zero faults or higher
degrees of fault tolerance, as discussed in \cite{gupta2016power}.
Authors in \cite{gupta2016power} propose a fault and power-aware
reliable resource allocation scheme that emphasizes the failure of
request occurs due to fluctuation in power consumption by the
requests. However, calculating the power consumption while scheduling both data and compute intensive requests is a cumbersome task. 

Handling faults that occur during the execution of real-time
tasks in cloud environment requires an efficient fault tolerant
mechanism, as demonstrated in \cite{wang2015festal}.
Though the proposed elastic resource provisioning mechanism based
on primary-backup model improves the resource utilization in the
context of fault tolerant, the faults that occur at the physical
server level cannot be handled by the proposed scheduling
algorithms. Authors in \cite{7976777} have proposed
similar fault tolerant scientific work-flow scheduling algorithm
considering the spot and on-demand instances on the cloud.

\section{Problem formulation }
\label{sec:probFormulation}

Unlike the applications running in conventional personal
computers, applications that are deployed in the cloud computing
environment and are accessed remotely through the web are known as
cloud applications. These applications are often collections of
multiple dependent or independent components known as cloud
components.
In the proposed scenario, each component can either be in an active or inactive
state. The active state of a component refers to a state in which
the cloud component is engaged in providing services to the cloud
user directly or indirectly, which is inactive otherwise. At any
given time $t$, a component is said to be an active component if the
same component is in the active state. Let, $n$ number of
active components from a total of $N$ number of components, the relationship between $n$ and $N$ can be 
written as $1 \leq n \leq N$.

\begin{figure}[h!]
    \centering
    \includegraphics[width=80mm]{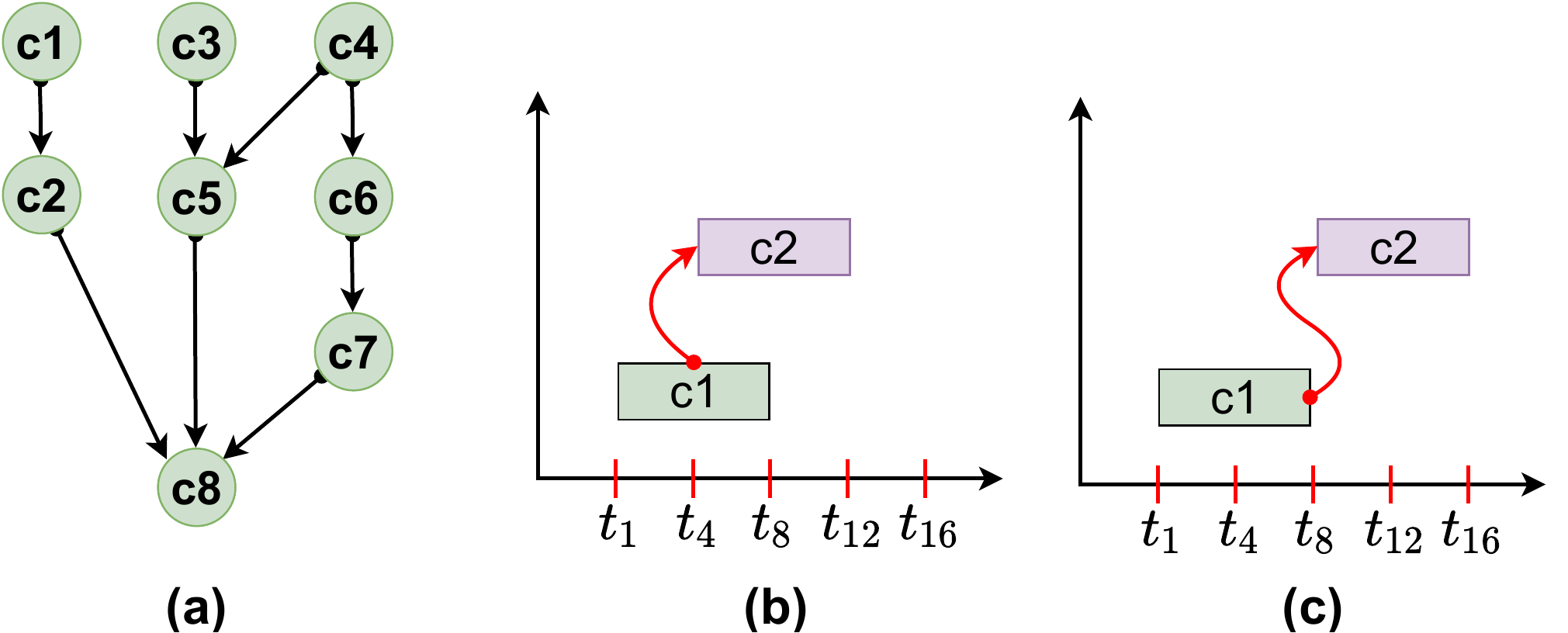}
    \caption{An example of DAG}
    \vspace{-5mm}
    \label{fig:DAGExample}
\end{figure}

A component is said to be dependent if it requires the
output or the intermediate result of another component.
A Directed Acyclic Graph (DAG) is used as a mathematical tool to
represent the dependencies among the active components. One vertex
in the DAG represents one active component. An edge represents the
dependency of one component over another one. For example, Figure
\ref{fig:DAGExample}(a) shows the dependencies of eight
components, where component $c2$ depends on the output of the
intermediate results of the component $c1$, and $c5$ requires the
output or intermediate results  of component $c3$ and $c4$ and so
on.

Based on the dependencies, a dependent component cannot start its
execution before the execution of the preceding component.
Figure \ref{fig:DAGExample}(b) and \ref{fig:DAGExample}(c)
illustrate this situation. In Figure \ref{fig:DAGExample}(b),
component $c2$ may start its execution after receiving the
intermediate result from component $c1$, whereas, in Figure
\ref{fig:DAGExample}(c), component $c2$ starts its execution after
receiving the final output from component $c1$. However, it is
assumed that a dependent component may finish its execution
before or after the execution of the preceding component. As
shown in Figure \ref{fig:DAGExample}, component $c2$ may finish its execution
before component $c1$ finishes its execution. It is also
assumed that a vertex cannot be isolated. In other words,
each component must be connected to at least one other component.
Among active components, multiple components can be executed
concurrently at any given instant of time. For example, component
$c2$, $c5$, and $c6$ can be scheduled to run concurrently as no
dependency exists among themselves.

Multiple cloud applications are hosted by cloud service
providers, where a vast number of cloud components are involved in
providing the service to the cloud users. The \emph{cloud
application} is considered as the input in this research article.
Based on the functionality, cloud components are compared and
different weights can be assigned to each component in order to
calculate their importance level. For example, in ATM withdrawal
system, different components are responsible for different
operations such as components for cash dispatching, receipt
printing, communicating remote database, etc. In such a scenario, the
importance level for the component responsible for remote database
communication is higher than the component that is responsible for
printing the withdrawal receipt. The component with higher
importance level, i.e., higher weight is considered as the 
\emph{most significant component}. Formally, a component is said
to be the most significant cloud component, if it is sufficiently
important to be worthy of providing the fault tolerant service.
The most significant cloud components are determined
programmatically.

In this work, mainly two aspects of fault tolerant services are
focused. Firstly, providing an efficient mechanism to rank components thereby 
determining the most significant cloud components and secondly, apply
a fault tolerant mechanism to the cloud components based on their
ranks so that the dedicated resources required to provide
fault tolerant service can be minimized. Let us take an example to
illustrate the latter aspect. Let, $A$, $B$, and $C$ be the three
components that are running concurrently with a maximum degree of
fault tolerance of $4$. The degree of fault tolerance indicates the
number of faults that can be tolerated. We assume that one VM is
used to host exactly one component. Considering the number of
components and the maximum degree of fault tolerance, we can
conclude that a total of $15$ VMs are required to
run $A$, $B$, and $C$ and their corresponding replicated components
concurrently. Let the importance level of component $A$ be ranked
as "High", component $B$ be as "Medium", and $C$ be ranked as
"Low". It is this importance of a component that translates to its rank value in our proposed algorithm.
With such knowledge
regarding the importance of each component and the corresponding
failure probability, the resources can be distributed unevenly
among all components for fault tolerant service. $6$ VMs can be
assigned to component $A$, $4$ VMs can be assigned to component
$B$, and $2$ VMs can be assigned to component $C$ for fault
tolerant service. As a result, a total of $12$ VMs can be used to
provide the fault tolerant service. Further, the degree of fault
tolerant can be increased for the components with higher
importance level. As discussed in this example, the number of
faults of component $A$ that can be tolerated is increased to
$6$.

The above example illustrates the goal of this paper, where a fault tolerant strategy needs to be developed that should rank the cloud components and employ a component-specific fault tolerant strategy based on the rank. The following section discussed in detail the procedure to rank the components and the fault tolerant strategy.

\section{The fault tolerant strategy}\label{sec:sol}
\label{sec:solution}
We believe that providing fault tolerant service to 
cloud application that consists of huge number of cloud components
is resource-intensive and hence inefficient in terms of monetary
cost. In order to address this issue, we propose a novel Rank
based Resource-aware Fault Tolerant (RRFT) strategy, which
provides fault tolerant services to selected significant
components instead of the whole application. The proposed RRFT
solution has two stages. Firstly, determining the most significant
component and secondly, providing the fault tolerant service to
each component based on their rank value, which indicates the
importance of the component.

\subsection{Determination of most significant component} \label{sec:sol:determine_sigCompont}
Here, the components of a cloud application are represented as a
DAG with no isolated vertex. In other words, at
any given time, each component must be connected to at least one
other component.
We use the group relationship model as a mathematical tool
followed by the probabilistic approach in order to find the most
significant cloud components of an application.
The goal of using group relationship model is to derive the
dependency of one active component on all others that belong to
the same cloud application. Let, $A$ be the cloud application to
which the fault tolerant service needs to be applied. $G=(C,E)$ is the
DAG of the set of components $C$ that belongs to cloud application $A$. $E$ represents the dependencies among
the components present in $C$. The set $C$ consists of $N$ components, represented by $c_i, 1\leq i \leq N$. Let,
$G_t^a=(C_t^a, E_t^a)$ be the DAG of set of active cloud
components $C_t^a$ and their dependencies $E_t^a$ of the
application $A$ at time $t$. $C_t^a = \{\hat{c}_1, \hat{c}_2,
\dots, \hat{c}_n\}$ consists of $n$ number of active components.
The relation between the sets $C_t^a$ and $C$ can be written as
$C_t^a \subseteq C$ at any given time $t$.

The DAG $G_t^a$ can be transformed to the adjacency matrix
$\tau_t$ as given below.

\begin{equation}
\tau_t = [ v_{ij} ], \quad 1\le i \le n, \ 1\le j \le n
\end{equation}
Where,
\[
v_{ij} = \left\{
\begin{array}{ll}
1 & \text{if the component }j\text{ depends on component } i;\\
0 & \text{if there is no dependency of component } i \text{ on }j;\\
0 & \text{if } i=j
\end{array} \right.
\]

The adjacency matrix is derived by looking at the edge set $E_t^a$
that contains the value ${1,0}$. The adjacency matrix is used to
calculate the dependency of component $i$ over component $j$,
where, $i \neq j, \forall j \in C_t^a$. The dependencies that are
present in edge set $E_t^a$ are called one-hop dependency. In
order to derive the multi-hop dependencies among components, the
distance matrix $\tau_t^d$ is calculated from the adjacency matrix
$\tau_t$ as derived below.

\begin{equation}\label{eq:distMatrix}
\tau_t^d = \begin{bmatrix}
d_{11} & d_{12} & \dots  & d_{1n} \\
d_{21} & d_{22} & \dots  & d_{2n} \\
\vdots & \vdots & \ddots & \vdots \\
d_{n1} & d_{n2} & \dots  & d_{nn}
\end{bmatrix}
\end{equation}


Where,
\begin{flalign}
\nonumber d_{ij} = \left\{
\begin{array}{ll}
d_{ij} & \text{minimum number of intermediate }\\
    & \text{components or minimum number of}\\
    & \text{ hops from component } i \text{ to } j.\\
0 & \text{If there is no path from component } i \text{ to }j.\\
0 & \text{If } i=j.
\end{array} \right. \\
\nonumber for \quad 1\le i \le n, \ 1\le j \le n
\end{flalign}

The significant value of an active component $\hat{c}_j, 1\leq j\leq n,$ can be derived by calculating the column sum of distance matrix $\tau_t^d$. Mathematically,

\begin{equation}\label{eq:sigficantVal}
\Psi(\hat{c}_j) = \sum_{i=1}^{n}{d_{ij}}, d_{ij} \in \tau_t^d, 1\leq i\leq n, 1\leq j\leq n
\end{equation}

The component with minimum significant value $\Psi(\hat{c}_j)$ \rthree{is considered the most influential component}. This is due to the fact that \rthree{several components depend on the }output of such components. For instance, the components $c2, c5, c6, c7,$ and $c8$ depend on the output of the components $c1, c3,$ and $c4$ as shown in Figure \ref{fig:DAGExample}. Hence, the components $c1, c3,$ and $c4$ could be considered as the most influential components.
This indicates that The failure of such
component has an impact on a maximum number of other components and
therefore has a higher impact on the whole cloud application. The
significant value of an active component is further used in the
calculation of failure impact of the respective component. The
\emph{failure impact} of one component $F(\hat{c}_i)$ is
calculated as the sum of the significant values of the components
$\hat{c}_j, 1\leq j\leq n, i\neq j$ that satisfies the conditions
$d_{ij}>0$. Mathematically,
\begin{equation} \label{eq:failureImpact}
F(\hat{c}_i) = \sum_{j=1}^{n}{\Psi(\hat{c}_j)}, \quad i\neq j, d_{ij}>0, 1\leq i\leq n
\end{equation}

Along with the failure impact, the historical performance is also
taken into account to calculate the \emph{accumulated failure
impact} of a component, $\ddot{F}(\hat{c}_i)$. The \emph{failure
rate} of a component is represented as $\lambda(\hat{c}_i),
\forall \hat{c}_i \in C_t^a$. The value of failure rate and
failure impact as calculated in Equation \ref{eq:failureImpact},
is multiplied in order to obtain the value of
$\ddot{F}(\hat{c}_i)$, as in Equation \ref{eq:AcfailureImpact}.

 \begin{equation} \label{eq:AcfailureImpact}
 \ddot{F}(\hat{c}_i) = \lambda(\hat{c}_i) * F(\hat{c}_i), \quad \forall \hat{c}_i\in C_t^a
 \end{equation}

Furthermore, the information regarding the failure probability
$P(\hat{c}_i)$ of a component $\hat{c}_i, \forall \hat{c}_i \in
C_t^a$ is essential in order to determine the most significant
component. It is assumed that the component $\hat{c}_i$ will
remain active for $h_i$ time units and failure of the components
follow the Poisson distribution \cite{anderson1976reliability}. Considering the aforementioned
information, the failure probability of a component can be
calculated as follows.

\begin{equation} \label{eq:failureProb}
P(\hat{c}_i) = \frac{\lambda(\hat{c}_i)*h_i}{e^{\lambda(\hat{c}_i)*h_i}}
\end{equation}

Here, it is assumed that $h_i$ is known. However, the average of $h_i$
can be taken into account, if the active time duration $h_i$ of
component $\hat{c}_i$ is unknown. The average active time duration
can be calculated by taking the ratio of total active time and
number of time the component $\hat{c}_i$ becomes active from
inactive state. Taking the value of accumulated failure impact as
given in Equation \ref{eq:AcfailureImpact}, and failure
probability as derived in Equation \ref{eq:failureProb}, the
\emph{most significant value} of each component is calculated as
derived below.
\begin{equation} \label{eq:mostSigValue}
 \Omega(\hat{c}_i) = \ddot{F}(\hat{c}_i) * P(\hat{c}_i), \quad \forall \hat{c}_i\in C_t^a
\end{equation}

As the complexity of developing cloud application increases, the complexity of 
monitoring the behavior of each component's execution
and providing the fault tolerant services also increases. The most significant value of a component is not enough
to determine whether the component should be given higher priority
while providing the fault tolerant service. The most significant
value $\Omega(\hat{c}_i)$ depends on various factors such as
 the number of times the component failed over its active time
duration, number of dependent components, etc. However, the
calculation of $\Omega(\hat{c}_i)$ does not consider the fact that a
single failure of any component may lead to the failure of the
whole cloud application.

In order to address this issue, the conventional joint probability
distribution approach is applied. The historical information
regarding the failure of the cloud application due to failure of a
specific component can be obtained from the performance history of
each component and the cloud application. Let, $f_i$ be the total
number of times component $\hat{c}_i$ failed. Further, let $f_i^s$
be the total number of times the whole cloud application is failed
due to the failure of component $\hat{c}_i$. Using the value of
$f_i$ and $f_i^s$, the mean application failure due to the failure of
component $\hat{c}_i$, $\lambda_i^s$, can be derived as follows.

\begin{equation}\label{eq:meanFalProbAppl}
\lambda_i^s = \frac{f_i^s}{f_i}, \quad 1\leq i\leq n
\end{equation}

Using Equation \ref{eq:meanFalProbAppl} and
Equation \ref{eq:failureProb}, the cloud application failure
probability due to the failure of component $\hat{c}_i$,
$\ddot{P}(\hat{c}_i|A)$ can be derived as 

\begin{equation} \label{eq:falImpOnAppl:1}
\ddot{P}(\hat{c}_i|A) = P(\hat{c}_i)  \left[ \frac{1}{x!} (\lambda_i^s h_i)^x e^{-\lambda_i^s h_i}\right]
\end{equation}

where, $h_i$ is the time duration of the active component $\hat{c}_i$. It is assumed that the failure of the components follows Poisson distribution. By putting the value of $x=1$ and the value
of $P(\hat{c}_i)$ in Equation \ref{eq:failureProb}, the Equation
\ref{eq:falImpOnAppl:1} can be re-written as

\begin{align} \label{eq:falImpOnAppl:2}
\ddot{P}(\hat{c}_i|A) = \left[ \frac{\lambda(\hat{c}_i) h_i}{e^{\lambda(\hat{c}_i) h_i}} \right] *
\left[ \frac{\lambda_i^s h_i}{e^{\lambda_i^s h_i}} \right] \nonumber \\
= \frac{(h_i)^2 \lambda_i^s \lambda(\hat{c}_i)}{e^{h_i(\lambda(\hat{c}_i) + \lambda_i^s)}}
\end{align}


\begin{table}[]
	\centering
	\begin{tabular}{ll}
		\hline
		\rowcolor[HTML]{CBCEFB} 
		\multicolumn{1}{|l|}{\cellcolor[HTML]{9698ED} $\Omega(\hat{c}_i)$ } & \multicolumn{1}{l|}{\cellcolor[HTML]{9698ED}  $\ddot{P}(\hat{c}_i|A)$ } \\ \hline
		\rowcolor[HTML]{C0C0C0} 
		\multicolumn{1}{|l|}{\cellcolor[HTML]{C0C0C0} $\hat{c}_2 = 0.115$ } & \multicolumn{1}{l|}{\cellcolor[HTML]{C0C0C0} $\hat{c}_3 = 0.922$ } \\ \hline
		\rowcolor[HTML]{EFEFEF} 
		\multicolumn{1}{|l|}{\cellcolor[HTML]{EFEFEF} $\hat{c}_4 = 0.0674$ } & \multicolumn{1}{l|}{\cellcolor[HTML]{EFEFEF} $\hat{c}_2 = 0.617$ } \\ \hline
		\rowcolor[HTML]{C0C0C0} 
		\multicolumn{1}{|l|}{\cellcolor[HTML]{C0C0C0} $\hat{c}_1 = 0.023$ } & \multicolumn{1}{l|}{\cellcolor[HTML]{C0C0C0} $\hat{c}_4 = 0.514$ } \\ \hline
		\rowcolor[HTML]{EFEFEF} 
		\multicolumn{1}{|l|}{\cellcolor[HTML]{EFEFEF} $\hat{c}_3 = 0.008$ } & \multicolumn{1}{l|}{\cellcolor[HTML]{EFEFEF} $\hat{c}_1 = 0.455$ } \\ \hline
		\rowcolor[HTML]{FFFFFF} 
		\multicolumn{2}{c}{\cellcolor[HTML]{FFFFFF}\textbf{(a)}}                           
	\end{tabular}
\quad
	\begin{tabular}{ll}
	\hline
	\rowcolor[HTML]{CBCEFB} 
	\multicolumn{1}{|l|}{\cellcolor[HTML]{9698ED}Rank ($\beta_i$)} & \multicolumn{1}{l|}{\cellcolor[HTML]{9698ED}Component ($\hat{c}_i$)} \\ \hline
	\rowcolor[HTML]{C0C0C0} 
	\multicolumn{1}{|l|}{\cellcolor[HTML]{C0C0C0} $\beta_2 = \beta_3 = 1$ } & \multicolumn{1}{l|}{\cellcolor[HTML]{C0C0C0} $\hat{c}_2, \hat{c}_3 $} \\ \hline
	\rowcolor[HTML]{EFEFEF} 
	\multicolumn{1}{|l|}{\cellcolor[HTML]{EFEFEF} $\beta_4 = 2$ } & \multicolumn{1}{l|}{\cellcolor[HTML]{EFEFEF} $\hat{c}_4$} \\ \hline
	\rowcolor[HTML]{C0C0C0} 
	\multicolumn{1}{|l|}{\cellcolor[HTML]{C0C0C0} $\beta_1 = 3$ } & \multicolumn{1}{l|}{\cellcolor[HTML]{C0C0C0} $\hat{c}_1$} \\ \hline 
	\rowcolor[HTML]{FFFFFF} 
	\multicolumn{2}{c}{\cellcolor[HTML]{FFFFFF} \textbf{(b)}}                           
	\end{tabular}
	\caption{An example of ranking the component}
    \label{table:rankingExample}
\end{table}

Considering the values calculated in Equation
\ref{eq:mostSigValue} and \ref{eq:falImpOnAppl:2}, components are
sorted individually. 
For the sake of better understanding on how $\Omega(\hat{c}_i)$ (Equation \ref{eq:mostSigValue}) and $\ddot{P}(\hat{c}_i|A)$ (Equation \ref{eq:falImpOnAppl:2}) values are used to rank all the components, an example is given in Table \ref{table:rankingExample}. It is assumed that the values given in Table \ref{table:rankingExample}(a) are already calculated.
The first column contains the value of $\Omega(\hat{c}_i)$, which represents the  most significant value of each active component. The second column contains the cloud application failure probability due to the failure of each component. The value in the first column ($\Omega(\hat{c}_i)$) and second column ($\ddot{P}(\hat{c}_i|A)$) are independent of each other. Based on the
value of $\Omega(\hat{c}_i)$ and $\ddot{P}(\hat{c}_i|A)$, four
components, $\hat{c}_1, \hat{c}_2, \hat{c}_3, \text{ and }
\hat{c}_4$ are sorted separately in descending order, as shown in
Table \ref{table:rankingExample}(a). In both sorted list, component
$\hat{c}_2, \text{ and } \hat{c}_3$ are at the top, and hence
assigned with the rank value $\beta_2 = \beta_3 = 1$, where the notations $\beta_2$ and $\beta_3$ represent the rank of component $\hat{c}_2$ and $\hat{c}_3$, respectively. Rank value $2$ is assigned to
the component $\hat{c}_4$ (i.e. $\beta_4 = 2$). Here, as component $\hat{c}_2$ is
already assigned with rank value $1$, another rank value cannot be
assigned. Similarly, rank values are assigned to other components,
as given in Table \ref{table:rankingExample}(b). It is observed
that the rank value of two components can be the same and must be
treated equally while providing the fault tolerant services.
Component with small rank value is treated as the higher priority
component and hence, maximum resource should be assigned to those
components in order to tolerate the higher degree of faults.

\subsection{Fault tolerant service: Hybrid k*} \label{sec:sol:faultTolerant}
\label{sec:solution:fault_service} With the given component list
and corresponding ranks, our goal is to provide fault tolerant
service consuming minimum
amount of resources to each component based on their rank. As discussed earlier, the service provided by
the CSP may encounter unexpected interruption due to the failure
of the component itself, the corresponding VM, or the failure of corresponding physical
machine. \emph{Replication} is one of the most popular fundamental
methods to handle the faults. In the case of failure, identical components are
scheduled to run either concurrently or sequentially. For example, two identical components, $\hat{r}_1^1$ and
$\hat{r}_1^2$ can be created from the active component $\hat{c}_1$. Those identical
components can be scheduled in two ways. Firstly, both identical
components can be scheduled on different VMs or the VM, where the active component $\hat{c}_1$ is scheduled. So, any fault
that occurs on any one component can be tolerated, as the results can be
obtained from other components, which are running concurrently.
Secondly, the components can be scheduled to run sequentially. If one
component is failed to provide the result, the other
identical component can be scheduled to run. In both parallel and
sequential execution, the requirement of resource and time is a
trade-off. The amount of resource required to run the identical
components in parallel order is more than the resource required to
schedule the identical components in sequential order. However,
the time required in parallel execution is less than that of the
sequential one.

While providing the fault tolerant service, two major factors need to be taken into account: (a) number of replicas or
identical instances of the primary component, (b) the order of
execution of the replica components. Considering those two factors, an efficient resource
aware fault tolerant mechanism, called \emph{Hybrid k*} is proposed in this paper. As the
name \emph{hybrid} suggests, the order of the execution of main or primary
component and its corresponding replica components vary among all
active components. For example, for one cloud application, the
execution order of one component, say $\hat{c}_1$, and its
replica components are parallel, whereas the execution order of
another component, says $\hat{c}_2$, and its replica components
are sequential. On the other hand, \emph{k*} indicates the
number of replicas $k$ for all active components at any particular
time varies.

Let, $\hat{r}_i^j$ be the $j^{th}$ replica component of 
$\hat{c}_i$. Furthermore, in general, the cloud applications are
associated with a hard deadline. To meet the deadline and to provide a higher degree of fault tolerance, the replica components must run
concurrently. The amount of resources can be minimized by scheduling
less significant replica components in sequential manner. It
is very challenging task to determine the order of execution of
the replica component. Let, $\Theta_i^p$ be the boolean variable
to indicate if the replicas of primary component $\hat{c}_i$ are
running concurrently.

\[
\Theta_i^p = \left\{
\begin{array}{ll}
1 & \text{if the replicas of primary component }\hat{c}_i \\
  & \text{are scheduled to run concurrently};\\
0 & \text{Otherwise }
\end{array} \right.
\]

Similarly, $\Theta_i^s$ is the boolean variable that indicates if
the replicas of the primary component $\hat{c}_i$ run
sequentially. Mathematically,

\[
\Theta_i^s = \left\{
\begin{array}{ll}
1 & \text{if the replicas of primary component}\hat{c}_i \\
& \text{are scheduled to run sequentially};\\
0 & \text{Otherwise}
\end{array} \right.
\]

Considering the goal to minimize the required resources, the
multi-criteria objective function for scheduling parallel
executions of the replicas would be as follows.

\begin{equation} \label{eq:objParallel}
\min P =\left[
\begin{array}{ll}
k_i * R_i^x \\
P(\hat{c}_i)^{k_i}
\end{array}
\right]
\end{equation}

Where, the term $k_i$ represents the number of replicas used to
provide the fault tolerant service to the primary component
$\hat{c}_i$. $R_i^x$ is the amount of resources of type $x$
required by the component $\hat{c}_i$. The resource type $x$ can
either be CPU or memory. Hence, the objectives in
Equation \ref{eq:objParallel} can be summarized as follows. Firstly, to minimize the amount
of resources required by the primary and corresponding
replica components. Secondly, to minimize the failure probability of the
component $\hat{c}_i$.

Similarly, considering the minimization of the time, the
multi-criteria objective function for scheduling sequential
executions of the replicas would be as follows.

\begin{equation} \label{eq:objSequential}
\min S =\left[
\begin{array}{ll}
k_i * \bigtriangleup_i \\
P(\hat{c}_i)^{k_i}
\end{array}
\right]
\end{equation}

Where, $\bigtriangleup_i$ represents the time required to start a
replica of component $\hat{c}_i$ on its failure. In sequential
execution, the total amount of resources required by the primary and all the replica components is equal to the amount
of resource required by only the primary component. By minimizing the value of $k_i$, the extra required
time can be minimized.

Two major parameters, i.e., permissible failure probability
$(\bigtriangledown)$ and threshold on number of replica components
$(\mu)$ are introduced to allow minimum fairness in providing
fault tolerant service among components. Permissible failure
probability of a component $\hat{c}_i$, represented by
$\bigtriangledown$, indicates the minimum failure probability allowed for any primary or replica component not to create
further replicas.

Combining the objective functions mentioned in Equation
\ref{eq:objParallel} and \ref{eq:objSequential}, the objective
function to minimize resource requirements and the time can be
derived as follows:

\textbf{Objective}\\
\begin{equation} \label{eq:objComponent}
\min \acute{d}_i = \Theta_i^p \left[
\begin{array}{l}
k_i * R_i^x \\
P(\hat{c}_i)^{k_i}
\end{array}
\right] + \Theta_i^s \left[
\begin{array}{l}
k_i * \bigtriangleup_i \\
P(\hat{c}_i)^{k_i}
\end{array}
\right],  \forall \hat{c}_i \in C_t^a
\end{equation}

The objective function mentioned in Equation \ref{eq:objComponent}
of all active components at time $t$ can  further be derived to
formulate the objective function of the cloud application $A$ as
follows:

\begin{equation} \label{eq:objectiveFun}
\min \breve{A} = \left[
\begin{array}{l}
\acute{d}_1 \quad \acute{d}_2 \quad  \acute{d}_3 \hdots \quad \acute{d}_n
\end{array}
\right]^T
\end{equation}
\textbf{Constraints:}\\
\begin{equation} \label{const1}
\text{if } \beta_i < \beta_j , \quad \Theta_i^p \leq \Theta_j^p,\text{ or }\Theta_i^s \geq \Theta_j^s
\end{equation}
\begin{equation} \label{const6}
\text{if } \beta_i = \beta_j , \quad \Theta_i^p = \Theta_j^p,\text{ or }\Theta_i^s = \Theta_j^s
\end{equation}
\begin{equation}\label{const2}
\Theta_i^p + \Theta_i^s = 1, 1\leq i \leq n
\end{equation}
\begin{equation}\label{const3}
P(\hat{c}_i)^{k_i} < \bigtriangledown
\end{equation}
\begin{equation}\label{const4}
k_i \geq \mu
\end{equation}
\begin{equation}\label{const5}
1\leq i \leq n, 1\leq j \leq n, i \ne j
\end{equation}

\begin{enumerate}
    \item Constraint (\ref{const1}) ensures that the replicas of a higher-ranked component must be scheduled to run concurrently in order to meet the deadline of the components. Further, the rank of the component, whose replicas are scheduled to run in sequential order, must be less than the rank of the component, whose replicas are scheduled to run in parallel order.
    \item Constraint (\ref{const6}) ensures that replicas of the components with the same ranks must be scheduled to run either in parallel or in sequential order. The replicas cannot be scheduled to execute in a different order.
    \item Constraint (\ref{const2}) ensures that the replicas of a component $\hat{c}_i$ are running in either parallel or sequential order.
    \item According to Constraint (\ref{const3}), the number of replicas must be determined in such a way that the failure probability with $k_i$ number of replicas must be less than the threshold value decided by the cloud service provider.
    \item We believe that component having failure probability less than $\bigtriangledown$ and with no replica may fail unexpectedly. To avoid such situations, we deduced constraint (\ref{const4}), which ensures that a minimum number of replicas decided by the objective function must be greater than or equal to the threshold value $\mu$. However, the value of $\mu$ should be decided by the cloud service provider by observing either the historical information or through a dedicated mechanism.
\end{enumerate}

In this paper, the primary components with higher rank are given highest priority to schedule their backup components in a parallel manner. Mathematically,
\begin{equation}
\Theta_i^p = 1, \forall \hat{c}_i, \beta_i > \beta_j, 1 \le j \le n, i \ne j
\end{equation}
For a sequence of components $\hat{c}_1, \hat{c}_2, \dots, \hat{c}_n$, let the rank of the components be $\beta_1 > \beta_2, \dots, \beta_n$. The rank of a component is directly proportional to $P(\hat{c}_i), \lambda_i^s, \ddot{F}(\hat{c}_i), \ddot{P}(\hat{c}_i|A)$ etc. 
This infers that failure of a component $\hat{c}_i$ with rank $\beta_i$ would have higher impact on the entire application as compared to the failure of a component $\hat{c}_{i-1}$ with rank $\beta_{i-1}$, if $\beta_i > \beta_{i-1}$. In order to reduce the failure impact of both the components $\hat{c}_i$ and $\hat{c}_{i-1}$, the backup components can be scheduled to run in a concurrent manner. However, this would bring additional resource overhead and the cost for providing fault tolerant service. In order to restrict the amount of resources to be used for fault tolerant service and the cost or in the case of limited resource availability, one of the components needs to be given higher priority to use the limited resources. In such scenarios, it is essential to give higher priority to component $\hat{c}_i$ over $\hat{c}_{i-1}$ as $\beta_i > \beta_{i-1}$. This approach can be extended and applied to all the component sequence given above. Hence, for the above component sequence, with given $\hat{c}_i$ and its rank $\beta_i$, if execution order is parallel, then for all the components $\hat{c}_1, \hat{c}_2, \dots \hat{c}_i$, the execution order of the corresponding back up components must be parallel.


\subsection{Component placement}\label{sec:sol:placement}
The data center consists of a huge number of physical machines
connected in fat tree topology. Upon arrival of a cloud
application, the cloud service provider carries out specific
procedures and selects the suitable physical machines for hosting
the incoming cloud applications. In the proposed scenario,
the placement of cloud component refers to the selection of
suitable physical machines (PMs) for hosting the cloud components.
Multiple cloud components are mostly involved in interacting with
each other to exchange the intermediate results among
them. The primary components are replicated to provide higher degree of fault tolerant services. 
The replica components are
placed either onto the same or different PMs. Upon failure of one
primary component, other primary components are connected to one of the
replica components of the failed primary component. Placement of
the replica onto a PM that is far away from other components leads
to performance degradation. Further, failure of a PM may have
a greater impact on the cloud application, if multiple primary and
replica components are placed onto the same PM. This motivates us
to propose a placement protocol to reduce the impact of a PM failure
on the specific cloud application.

In order to achieve the goal of reducing the impact of PM failure, the placement of components (primary and replicas) must follow the rules discussed below.

\textbf{Rule 1: \emph{One component on one PM.}}
\begin{equation}
    \sum_{i=1}^{n}{\kappa(S_j, \hat{c}_i)}, \quad \forall \hat{c}_i \in C_t^a
\end{equation}
Where $\kappa$  is the boolean variable indicating if the component $\hat{c}_i$ is scheduled to run in a PM $S_j$. Mathematically,
\[
 \kappa(S_j, \hat{c}_i) = \left\{
\begin{array}{ll}
1 & \text{if component }\hat{c}_i \text{ is placed onto PM }S_j;\\
0 & \text{Otherwise; }
\end{array} \right.
\]

According to the rule mentioned above, there is no restriction on allocating multiple components from one cloud application on single PM.
Failure at PM level leads to the failure of all virtual machines
running on that PM and therefore, it leads to the failure of multiple
primary components. Spreading the primary components among all
available PMs will reduce the impact of a single PM.

\textbf{Rule 2: \emph{Replica component must be in different PM than that of the PM of respective primary component.}}
\begin{equation}
\omega(\hat{r}_i^j) \neq \omega(\hat{c}_i), \quad 1\leq i \leq n, 1\leq j \leq k_i
\end{equation}
$\omega(\hat{c}_i)$ represents the PM, where the component
$\hat{c}_i$ is placed. The replica component $\hat{r}_i^j$ and its
corresponding primary component $\hat{c}_i$ cannot be scheduled to
run in the same PM. As discussed earlier, the failure of a PM leads
to the failure of all virtual machines running on that PM and
hence, the failure of all primary and replica components. The
replica component must be available to provide the uninterrupted
service in case of failure of primary components, which can only
be done by assigning different PMs for all the replicas of a
single primary component.

\textbf{Rule 3: \emph{No multiple replica from single primary component onto same PM.}}
\begin{equation}
\omega(\hat{r}_i^j) \neq \omega(\hat{r}_i^u), \quad 1\leq j \leq k_i, 1\leq u \leq k_i, j \neq u
\end{equation}
Similar to Rule 2, in order to reduce the impact of PM failure
onto single cloud application, the failure of single PM must have
the impact on utmost one component and hence, multiple replicas
cannot be scheduled onto single PM.

\textbf{Rule 4: \emph{Primary and the replica components must be
in the same pod.}} Pod in a data center generally refers to a
small container with multiple physical machines and one access
switch. To minimize the recovery time of a components'
execution from any failure, the backup component must be placed
with minimum distance from the location of primary component, but
in different PM.

\subsection{RRFT Algorithm} \label{sec:sol:algoRRFT}

Considering the components' rank as discussed in Section
\ref{sec:sol:determine_sigCompont}, \emph{k*} number of backup
components and the corresponding order of execution of the
components as discussed in Section \ref{sec:sol:faultTolerant},
and the placement of the components as discussed in Section
\ref{sec:sol:placement}, we propose Rank-based Resource aware
Fault Tolerant (RRFT) strategy for cloud application, which
analyzes the characteristics of each component and determines the
fault tolerance strategy. The details of the proposed RRFT is
presented in Algorithm \ref{algo:RRFT}.

\begin{algorithm}[t]
    \caption{Rank-based Resource aware Fault Tolerant algorithm}    \label{algo:RRFT}
    \KwIn{Set of $n$ active components in DAG form.}
    Calculate the distance matrix $\tau_t^d$ for all components using Equation \ref{eq:distMatrix}\; \label{algo:RRFT:calcDistMatrix}
    Calculate the failure impact $F(\hat{c}_i)$ for all components using Equation \ref{eq:failureImpact}\; \label{algo:RRFT:FailureImpact}

    Calculate accumulated Failure Impact $\ddot{F}(\hat{c}_i)$ using Equation \ref{eq:AcfailureImpact}\;

    Obtain the failure probability $P(\hat{c}_i)$ of all components \;

    Calculate the most significant value $\omega(\hat{c}_i)$ of all components using Equation \ref{eq:mostSigValue} \; \label{algo:RRFT:calcMSV}

    Calculate $\ddot{P}(\hat{c}_i|A)$ using Equation \ref{eq:falImpOnAppl:2} \; \label{algo:RRFT:calcSyFailProb}

    Rank the active components based on the value calculated in Step-\ref{algo:RRFT:calcMSV} and \ref{algo:RRFT:calcSyFailProb} \; \label{algo:RRFT:rankComponents}

    $CT = $ sort all active components based on their rank calculated in Step - \ref{algo:RRFT:rankComponents}\;\label{algo:RRFT:sortComponents}
    \tcc{Determine number of components and the execution order.}

    \ForEach{Component $c$ in $CT$}{ \label{algo:RRFT:determine_hybrid_k}
        Determine the number of backup components and their corresponding order of execution using Equation \ref{eq:objectiveFun} \;
    }
    \tcc{Placement of all active and backup components.}
    \ForEach{Component $c$ in $CT$}{ \label{algo:RRFT:compPlacement}
        Choose suitable PM by following rules mentioned in Section \ref{sec:sol:placement}\;
    }
\end{algorithm}

As discussed earlier, the cloud application consisting of a set of
active components is the input to the RRFT algorithm. Each component is analyzed
and is ranked based on the dependencies among each other, as given
in Line \ref{algo:RRFT:calcDistMatrix} to
\ref{algo:RRFT:rankComponents}. In order to rank the components,
the distance matrix of all the components is calculated as given
in Line \ref{algo:RRFT:calcDistMatrix}. The distance matrix is
further used to calculate the failure impact of all components as
given in Line \ref{algo:RRFT:FailureImpact}. The failure impact of
a component indicates the impact of one component failure onto the
other one. For example, the failure of the root component has
the highest impact onto all other components, whereas, the failure of
a leaf component at the bottom level of the DAG has no impact onto
other components. The failure impact and the failure rate of the
component are used to calculate the accumulated failure impact of
the components as given in Equation \ref{eq:AcfailureImpact},
which is further used to calculate the most significant value of
each component, as given in Line \ref{algo:RRFT:calcMSV}. In order
to calculate the most significant value, the failure probability
is used. It is assumed that the failure rate of the components is
known to the cloud provider from the historical data. It is
observed that the failure of a single component may lead to the
failure of the entire cloud application. Such historical
information can be used to predict the probability of application
failure due to the component's failure, as calculated in
Line \ref{algo:RRFT:calcSyFailProb}. The components are ranked as
given in Line \ref{algo:RRFT:rankComponents} based on the value
calculated in Line \ref{algo:RRFT:calcMSV} and
\ref{algo:RRFT:calcSyFailProb}.

A component's rank indicates its importance over
other components. Following the ranking of the component, the
number of backup components required by each primary
component is determined, as given in Line
\ref{algo:RRFT:determine_hybrid_k}. Besides, determining the
number of components, the order of execution is also determined.
The order of execution of the backup components can either be
parallel or sequential. The backups of the components with higher
rank are scheduled to run in parallel, whereas the order of
execution of the backup components of the components with lower
rank is scheduled to run in sequential manner. It is also
essential to place the components onto suitable physical machines.
Placement refers to creating the VMs with the corresponding
components onto the suitable physical machines. The placement of
the backup components is determined in Line
\ref{algo:RRFT:compPlacement}.



\section{Total number of replica optimization}
\label{sec:performance} 
\begin{figure*}[!h]
    \centering
    \includegraphics[width=130mm]{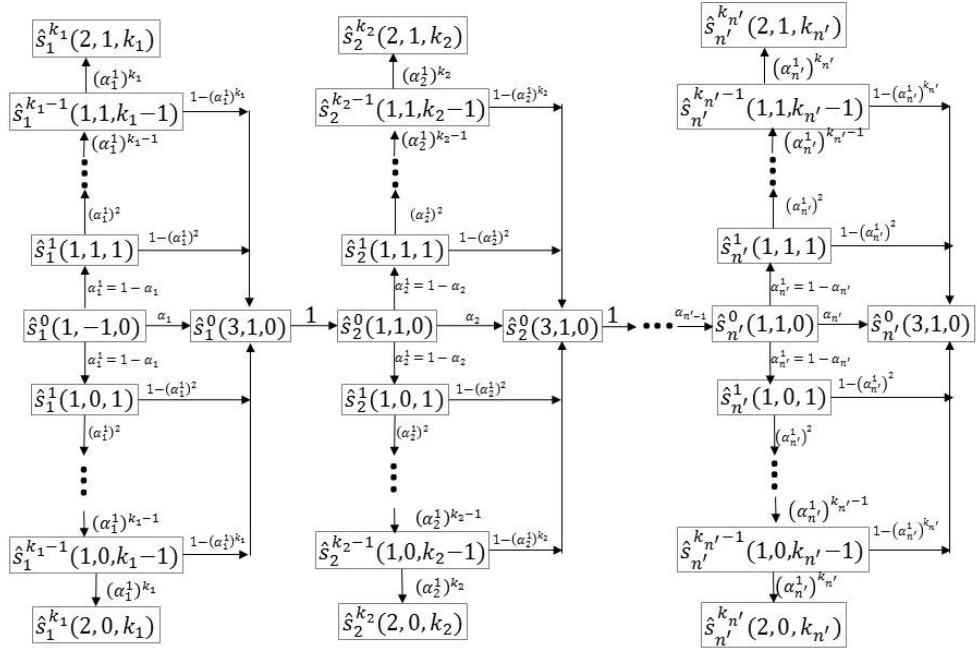}
    \caption{Proposed Markov decision process model}
    \vspace{-5mm}
    \label{fig:MDP_model}
\end{figure*}

In order to determine the number of
replicas for each primary active component as discussed in Section
\ref{sec:solution:fault_service}, we formulate the problem using
discrete-time three-dimensional Markov Decision Process (MDP)
model as depicted in Figure \ref{fig:MDP_model}.
MDP models are useful when the whole system is a discrete-time stochastic control process. It is also applied in addressing different research challenges of cloud computing such as resource management \cite{9112332}, QoS \cite{LIANG2021101991}, resource and service failure management \cite{li2019energy}, etc. As we need to make the decision on the type and number of replicas for each component, which is dynamic in nature due to the respective failure probability, it is highly essential to model the problem with a tool that can take the fully observable states of this controlled systems into consideration.
 
Based on the failure probability of a component, the number of
states is decided. A state represents either a primary or a
backup component. A component with higher failure probability may
have a larger number of states that may run either in parallel or
sequential manner. The MDP can be represented as 3-tuple
$\{\hat{S}, U, \hat{\alpha}\}$, where $\hat{S}$ represents the
state space, $U$ represents the action set that needs to be
performed in every state, and $\hat{\alpha}$ represents the set of
transition probabilities between two adjacent states.

\subsection{State space}
As discussed earlier, the state of a component represents the execution order of either a 
primary component or a backup component, which may run either in
parallel or sequential order. In the proposed MDP model, the state
of a component $\hat{c}_i, 1\leq i \leq n$ is represented as
$\hat{s}_i^j(\ddot{c}_i^j, \ddot{o}_i^j, \ddot{b}_i^j)$. $\hat{S}$
is the set of states for all primary and backup components. A
state captures three essential information. Those are (a) the
status of the component, (b) order of execution of corresponding
components, and (c) number of backup components.
Here, the value of superscript $j$ is dynamic and the maximum
value of $j$ for each component may vary depending on the failure
probability of the component. Value of $j=0$ represents that the
state is for the primary component. However, value of $j \geq 1$
indicates that the state is for a backup component. For example,
state $\hat{s}_1^0$ represents the primary component of component
$\hat{c}_1$, whereas the state $\hat{s}_1^2$ represents the
$2^{nd}$ backup component of the component $\hat{c}_1$. The status
of an active component $\hat{c}_i$ could be running, rejected or
finished, which is represented as $\ddot{c}_i^j$. Mathematically,

\begin{equation}
\ddot{c}_i^j = \left\{
\begin{array}{ll}
1 & \text{If component }\hat{c}_i \text{ is in running state.};\\
2 & \text{If component }\hat{c}_i \text{ is rejected.};\\
3 & \text{If component }\hat{c}_i \text{ is finished its execution.};
\end{array} \right.
\end{equation}

The order of execution of a replica component is represented as
$\ddot{o}_i^j$ of the active component $\hat{c}_i$.
Mathematically, the value of $\ddot{o}_i^j$ can be written as
follows.

\begin{equation}
\ddot{o}_i^j = \left\{
\begin{array}{ll}
0 & \text{If the replica components are scheduled to run }\\
   &  \text{concurrently.}\\
1 & \text{If the replica components are scheduled to run }\\
   &  \text{sequentially.}\\
-1 & \text{Otherwise.}
\end{array} \right.
\end{equation}

The order of execution $\ddot{o}_i^j$ for the order component is
-1. For example, the value of $\ddot{o}_i^j$ of the components
$c_1, c3,$ and $c4$ in Figure \ref{fig:DAGExample} is $-1$. This
indicates that such primary active components need to be scheduled
at the very beginning and do not depend on any other components.
The replica components can be scheduled to run concurrently,
represented by 0, or sequentially, represented by 1, as shown in
Figure ~\ref{fig:MDP_model}. However, as discussed earlier, the
execution order of replica components of the primary components
must follow the Constraint \ref{const1}. In other words, replicas
of higher-ranked primary component must be scheduled to execute
concurrently and the replicas of lower-ranked primary component must
be scheduled to run sequentially.

Eventually, the number of replica components is considered as the
third dimension of a state. The number of replicas of the
component $\hat{c}_i$ is represented as $\ddot{b}_i^j$. For
example, the state $\hat{s}_2^2(1,1,2 )$ indicates that two
replicas of the component $\hat{c}_2$ are scheduled to run
sequentially.

\subsection{Action space}
The set of actions that is available at every state is represented
as $U$. Any one of the three actions can be taken at each state,
such as (a) \textit{Create new backup} denoted as $u_1$, (b)
\textit{Reject the component}, denoted as $u_2$, and (c)
\textit{Execute the component}, denoted as $u_3$. Based on the
failure probability of the components, actions are applied to the
states. For example, applying the action $u_3$ with the success
probability $\alpha_1$ onto the state $\hat{s}_1^0(1,-1,0)$, the
resultant state would be $\hat{S}_1^0(3,1,0)$. Similarly, applying
the action $u_1$ onto the state $\hat{S}_1^0(1,-1,0)$, the
resultant state would be $\hat{S}_1^1(1, 0, 1)$.

\subsection{State transition probability}
It is assumed that $\alpha_i$ be the success probability of the
primary active component $\hat{c}_i, 1\leq i \leq n$ with no
corresponding replica components. Mathematically, the transition
probability from the state $\hat{s}_i^0(1, \ddot{o}_i^0, 0 )$ to
$\hat{s}_i^0(3, 1, 0 )$ with the action $u_3$ is $\alpha_i$.

\begin{equation}
\label{eq:probability_alpha_i}
 Pr(\hat{s}_i^0(3, 1, 0 ) | \hat{s}_i^0(1, \ddot{o}_i^0, 0 ), u_3) = \alpha_i, \quad 1\leq i \leq n
\end{equation}

From Equation \ref{eq:probability_alpha_i}, it can be derived that
the failure probability of the component $\hat{c}_i$. In order to
reduce the failure probability of a component, multiple replicas
can be scheduled to run either in parallel or sequential order.
Hence, the probability of creating the initial replica can be
calculated as $1- \alpha_i$. Mathematically,

\begin{equation}
\label{eq:probability_1-alpha_i}
\alpha_i^1=Pr(\hat{s}_i^1(1, \ddot{o}_i, 1 ) | \hat{s}_i^0(1, \ddot{o}_i^0, 0 ), u_1) = 1-\alpha_i, \quad 1\leq i \leq n
\end{equation}

Here, $\alpha_i^1$ represents the transition probability from the
state $\hat{s}_i^0$ to the state $\hat{s}_i^1$. In general, the
notation $\alpha_i^j$ represents the transition probability from
the state $\hat{s}_i^{j-1}$ to the state $\hat{s}_i^j, 1 \le j$.
As discussed earlier in Section \ref{sec:solution:fault_service},
$\bigtriangledown$ is defined as the minimum permissible failure
probability. The number of replica components depend on the value
of $\bigtriangledown$. The condition to apply the action $u_1$
onto any state can be written as follows.

\begin{equation}
\label{eq:replica_create_condition}
Pr(\hat{s}_i^{j+1}(1,\ddot{o}_i^j, \ddot{b}_i^{j+1})| \hat{s}_i^j(1,\ddot{o}_i^j, \ddot{b}_i^j), u_1) > \bigtriangledown
\end{equation}
In other words, a new replica component (i.e. to state $\hat{s}_i^{j+1}(1,\ddot{o}_i^j, \ddot{b}_i^{j+1})$) needs to be created (from the state $\hat{s}_i^j(1,\ddot{o}_i^j, \ddot{b}_i^j)$) of the component $\hat{c}_i$, by applying the action $u_1$, if the failure probability of the component is greater than $\bigtriangledown$.

From Equation \ref{eq:probability_1-alpha_i} and
\ref{eq:replica_create_condition}, and with the given value of
$\ddot{o}_i^j$, the order of execution of the replicas of the
primary active component $\hat{c}_i$, the number of replica
components can be determined from the following relation.

\begin{equation}
\label{eq:relation_k_probThreshold}
(1-\alpha_i)^{k_i} \le \bigtriangledown
\end{equation}

Where, $k_i$ is the number of replica components of the primary
active component $\hat{c}_i$. Further, using Equation
\ref{eq:relation_k_probThreshold}, the value of $k_i$ can be
derived as follows.

\begin{gather}
(1- \alpha_i)^{k_i} \le \bigtriangledown \nonumber \\ \nonumber
\qquad \Longrightarrow k_i*log(1-\alpha_i) \le log(\bigtriangledown) \\
\qquad \Longrightarrow k_i = \lceil log_{(1-\alpha_i)}(\bigtriangledown)\rceil
\end{gather}

In the following subsection, we will present a numerical example to illustrate the workings of our proposed MDP.

\subsection{Markov model example}

\begin{figure}[h]
    \centering
    \includegraphics[width=88mm]{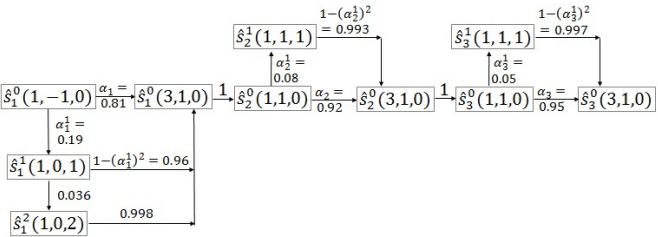}
    \caption{An example of proposed MDP model}
    \label{fig:MDP_model_example}
\end{figure}

In order to illustrate the proposed MDP model presented in Figure
\ref{fig:MDP_model}, let us assume a cloud application with three
components $\hat{c}_1, \hat{c}_2,$ and $\hat{c}_3$. Component
$\hat{c}_2$ and $\hat{c}_3$ depends on $\hat{c}_1$ and
$\hat{c}_2$, respectively. The initial state for component
$\hat{c}_1$ can be written as $\hat{s}_1^0(\ddot{c}_1^0=1,
\ddot{o}_1^0=-1, \ddot{b}_1^0=0)$. Similarly, for the component
$\hat{c}_2$ and $\hat{c}_3$, the states can be written as
$\hat{s}_2^0(1,1,0)$ and $\hat{s}_3^0(1,1,0)$, respectively. The
value of $\ddot{o}_2^0$ and $\ddot{o}_3^0$ for component
$\hat{c}_2$ and $\hat{c}_3$ is 1 as both components run
sequentially after the component $\hat{c}_1$. Let, the failure
probability of the component $\hat{c}_1, \hat{c}_2, $ and
$\hat{c}_3$ be $0.19, 0.08,$ and $0.05$, respectively, and the
value of minimum permissible failure probability
$\bigtriangledown$ be $0.007$.

For the above scenario, the MDP is presented in Figure
\ref{fig:MDP_model_example}. In Figure
\ref{fig:MDP_model_example}, the transition probability from the
state $\hat{s}_1^0(1,-1,0)$ to state $\hat{s}_1^1(1,0,1)$ is
$0.19$. State $\hat{s}_1^1(1,0,1)$ represents the first backup
component of the primary component $\hat{c}_1$ and is scheduled to
run in parallel. The failure probability of the first backup
component is calculated to be $0.036$, which is still greater than
the permissible failure probability $\bigtriangledown=0.007$. As a
result, the second backup component is created and is represented
as the state $\hat{s}_1^2(1,0,2)$. Since, the failure probability
of the state $\hat{s}_1^2(1,0,2)$ is calculated to be $0.002$,
which is less than $\bigtriangledown$, no further backup component
is created. Similarly, the failure probability of the component
$\hat{c}_2$ is $0.08$, which is greater than the value of
$\bigtriangledown$. In order to reduce the failure probability,
another backup component $\hat{s}_2^1(1,1,1)$ is created, which is
scheduled to run in a sequential manner. The state
$\hat{s}_3^0(3,1,0)$ represents that the component $\hat{c}_3$ has
finished its execution.

\subsection{Policy optimization}
Using the proposed Three-dimensional MDP model as discussed above,
we optimize the policy of determining the total number of replica
components for all primary active components. Mathematically,
\begin{equation}
Minimize\text{ }K = \sum_{i=1}^{n} k_i, \qquad 1 \le i \le n
\end{equation}

where, $K$ is the sum of number of replica components for all
primary active components. In order to optimize the value of $K$,
it is necessary to minimize the value of $k_i$ for each component.
However, the policy must follow the constraint mentioned in
Constraint \ref{const1}. In other words, replicas of the component
with the highest rank must be scheduled to run in parallel and the
replicas of the component with lower rank must be scheduled to
execute in sequential manner. This will support our belief that
most significant components must finish their execution with
higher degree of fault tolerant without violating the deadline. On
the other hand, replicas of the lower rank components are allowed
to take longer time in case of any failure as the failure of lower
rank components has a trivial impact on other components.

\section{Performance Evaluation}
\label{sec:Perf_Evaluation} In order to evaluate the performance
of the proposed RRFT strategy, the proposed component ranking and
placement scheme is implemented on MATLAB platform. The cloud
applications are generated randomly and are presented
in matrix array. The cloud applications with varied numbers of
components are generated using different probabilistic approaches
such as arrival of cloud application from users that follows the
Poisson distribution. We compare our component ranking method
against two popular ranking algorithms, i.e., FTCloud
\cite{zheng2012component} and ROCloud \cite{qiu2014reliability}.
In FTCloud, component invocation structure and invocation
frequencies parameters are used to rank the most significant
components. Another version of the algorithm combines the
application designer knowledge to find the critical and
non-critical component and the system structure to rank the cloud
components. On the other hand, the algorithm proposed in ROCloud
uses failure rate and the failure impact as the reliability
properties of the components. The extended version of ROCloud
considers the hybrid applications, which consists of components
that can be migrated to the public cloud and some components that
can be scheduled to run in private cloud.

Further, the proposed component placement algorithm is evaluated
and compared with random placement method. The cloud application
is presented in DAG, which indicates that the
adjacency matrix does not contain any circle. The failure rate of
the components follows Poisson distribution in the simulation. The
active time of the components is assumed to be known, which is
assigned randomly in the simulation. In order to reduce the impact
of a single failure of a physical machine, one PM is assigned to
host one component from one cloud application. The number of
components in each cloud application is randomly distributed,
ranging from $4$ through $16$. In such scenarios, the minimum
number of required PMs will range from $4$ through $20$. The
probability that two components are connected could range from
$0.5$ through $0.8$. The memory and CPU resource requirement of
the components follows the random distribution ranging from
$1000MB$ through $2000MB$ and from $1$ through $4$, respectively.
Similarly, the memory and CPU capacity of the PMs are distributed
randomly, ranging from $16000MB$ to $32000$ and $16$ to $32$,
respectively. Taking above-mentioned performance matrix, following
simulation results are obtained.

\subsection{Simulation Results}\label{sec:Perf_Evaluation:simResult}
\begin{figure}[h]
    \centering
    \includegraphics[width=85mm]{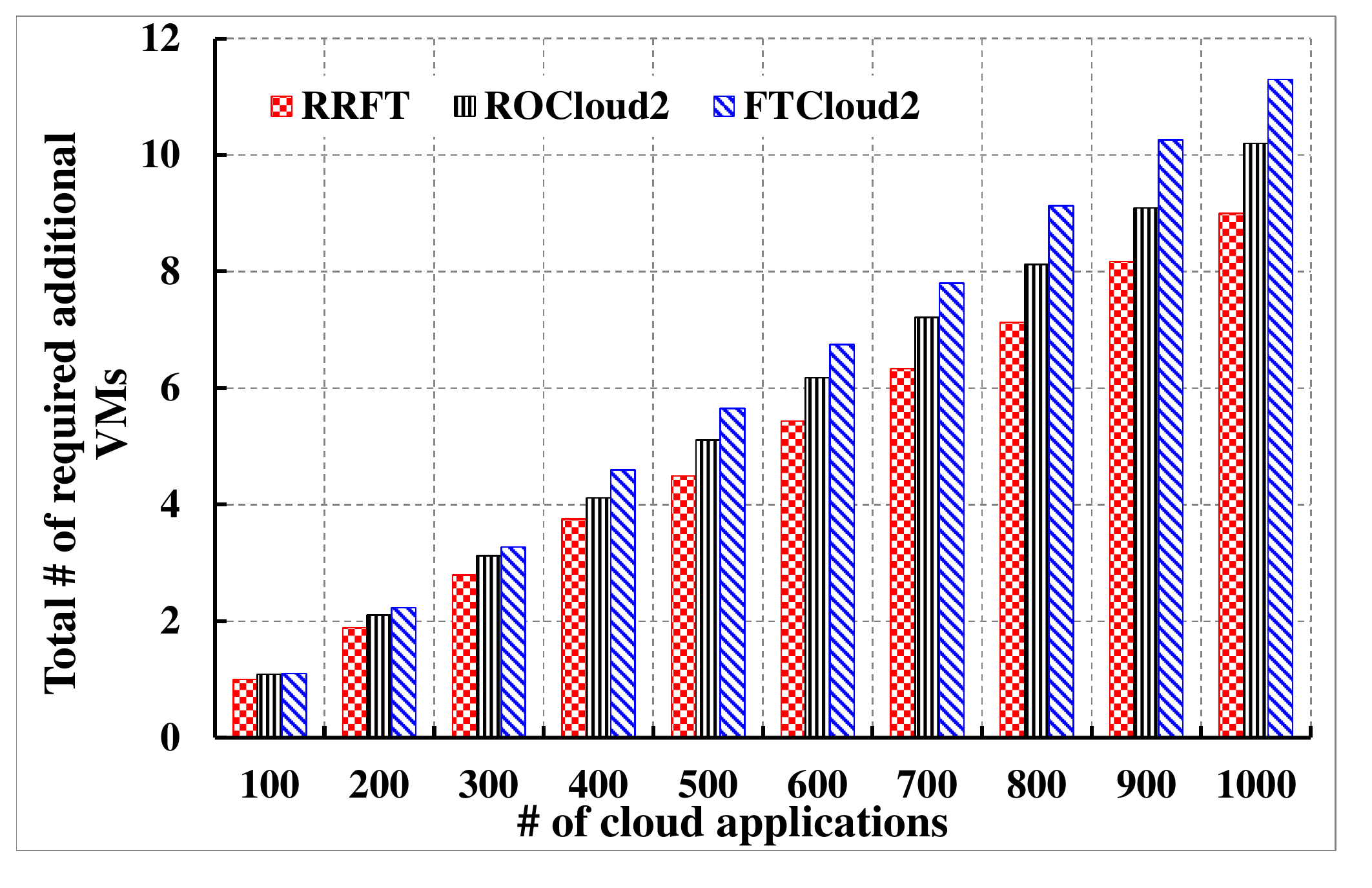}
    \vspace{-0mm}
    \caption{Number of required VMs.}
    \label{fig:sim:noofVMreq}
\end{figure}

\begin{figure}[h]
    \centering
    \includegraphics[width=75mm]{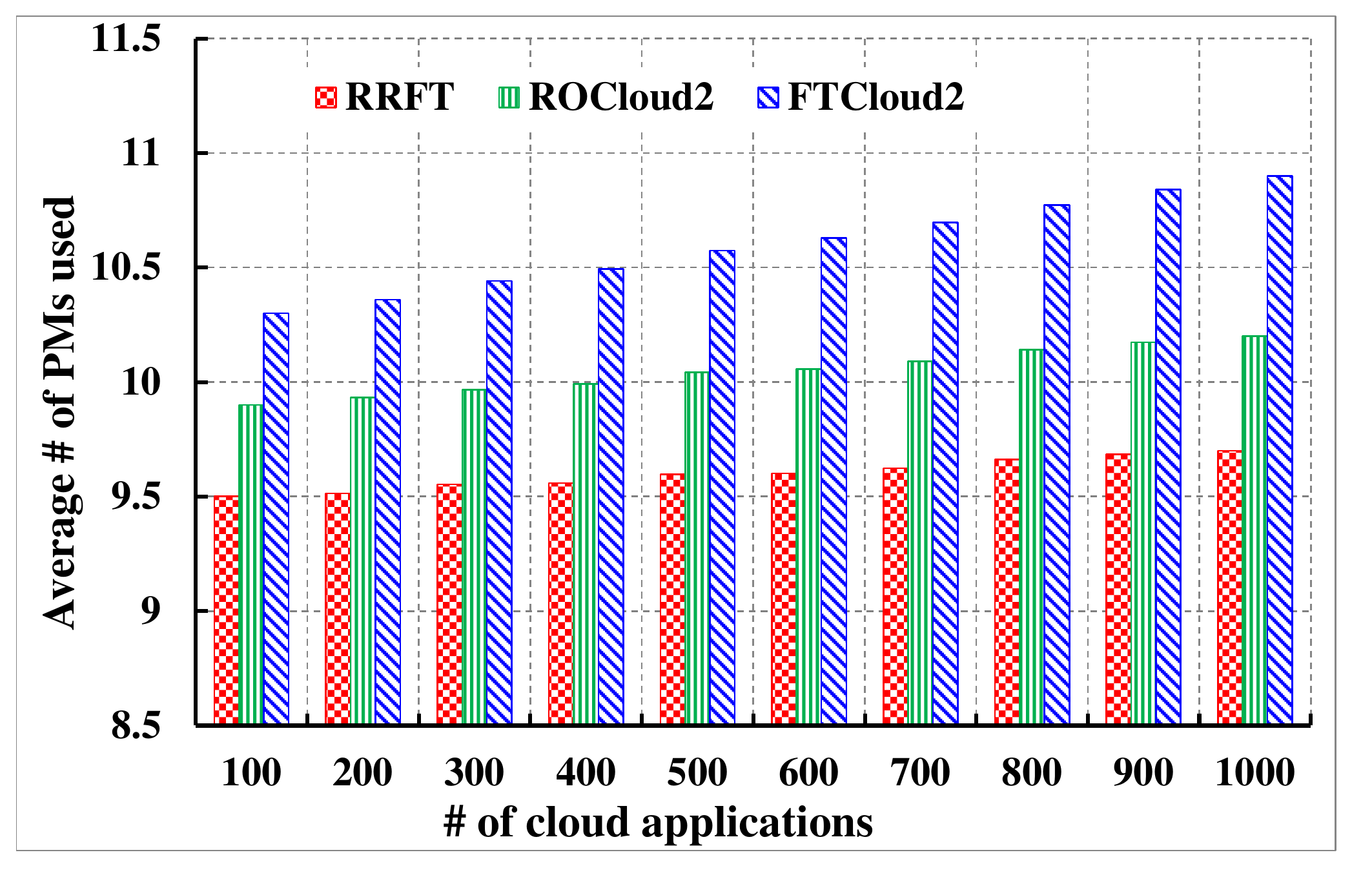}
    \vspace{-3mm}
    \caption{Number of required PMs.}
    \label{fig:sim:noofPMreq}
\end{figure}

The performance of the proposed scheme is compared with ROCloud2
and FTCloud2 algorithm, as discussed before. The comparison results
are shown in Figure \ref{fig:sim:noofVMreq} and
\ref{fig:sim:noofPMreq}. The results shown in Figure
\ref{fig:sim:noofVMreq} reveal the total number of VMs required
for a certain number of cloud applications. The number of cloud
application varies from $100$ to $1000$. It is observed from the
simulation result that the total number of VMs required for $100$
cloud applications is $996$, $1090$, and $1107$ for the proposed
RRFT algorithm, ROCloud2, and FTCloud2 algorithms, respectively.
However, the number increases to $9010, 10205,$ and $11317$ for
$1000$ numbers of cloud applications in case of RRFT, ROCloud2,
and FTCloud2 algorithms, respectively. This shows approximately a 10\% reduction in the required number of VMs when the number of cloud applications is 1000. It is expected that the RRFT algorithm would give similar result even after deploying more than 1000 cloud applications. In the simulation setup, the number of components per cloud application ranges from $4$ to $16$.

Similarly, the number of PMs used to host a certain number of cloud
applications is shown in Figure \ref{fig:sim:noofPMreq}. The value
along Y-axis shows the average number of PMs used to host certain
number of cloud applications. For example, the average number of
PMs used to execute $100$ numbers of cloud applications is $9.5$
in the case of proposed RRFT algorithm. However, in case of
ROCloud2 and FTCloud2 algorithm, the average number of PMs used is
$9.9$ and $10.3$, respectively, which is at least 4.2\% more than that of RRFT algorithm. The required number of PMs increases to $9.7, 10.2,$
and $10.9$ for $1000$ number of cloud applications while executing
the RRFT, ROCloud2, and FTCloud algorithm, respectively.

\begin{figure}[h]
    \centering
    \includegraphics[width=80mm]{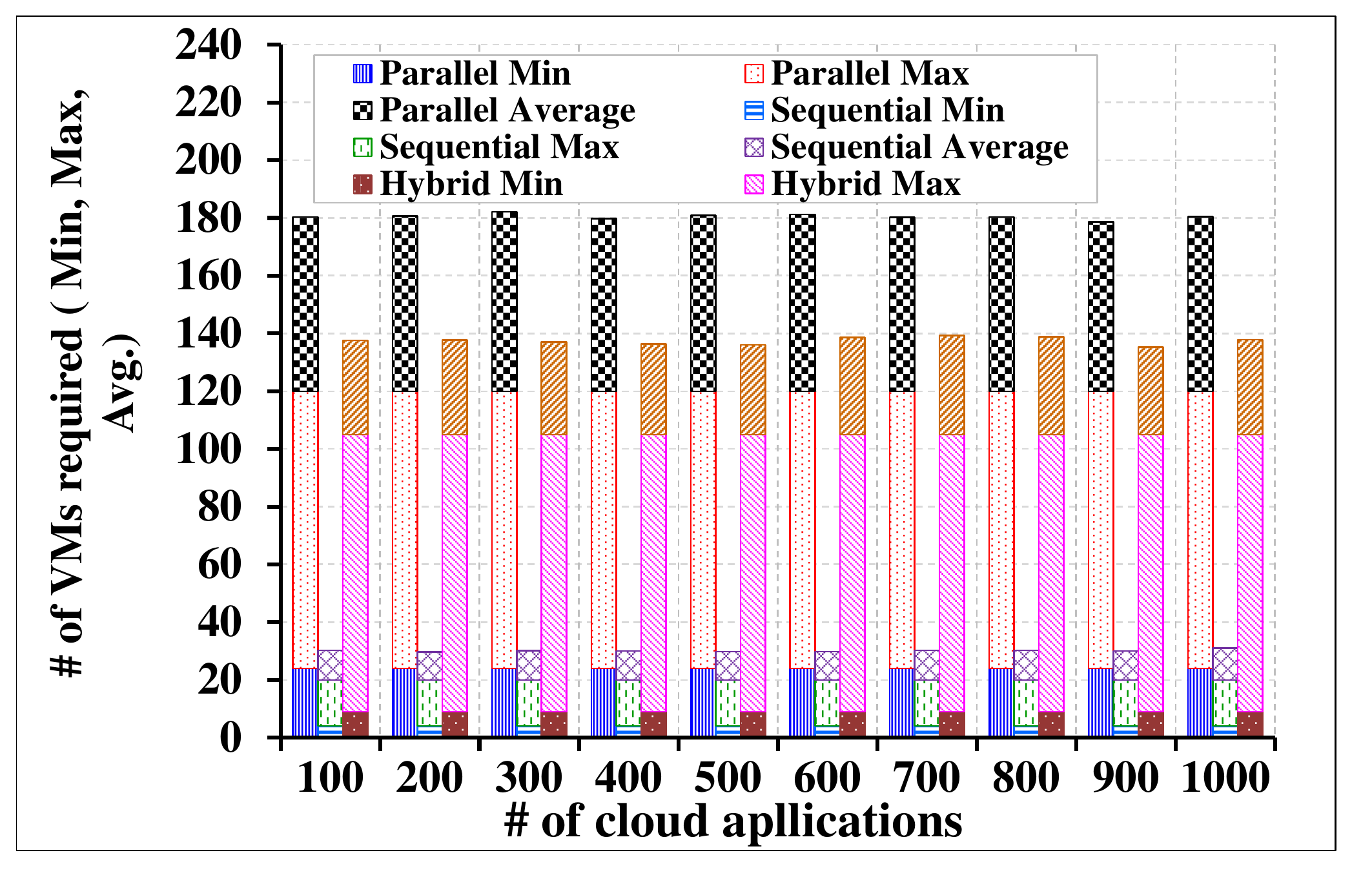}
    \vspace{-4mm}
    \caption{Number of VMs required under different order of execution of components.}
    \label{fig:sim:noofVMreq_ExecutionOrder}
\end{figure}

\begin{figure}[h]
    \centering
    \includegraphics[width=75mm]{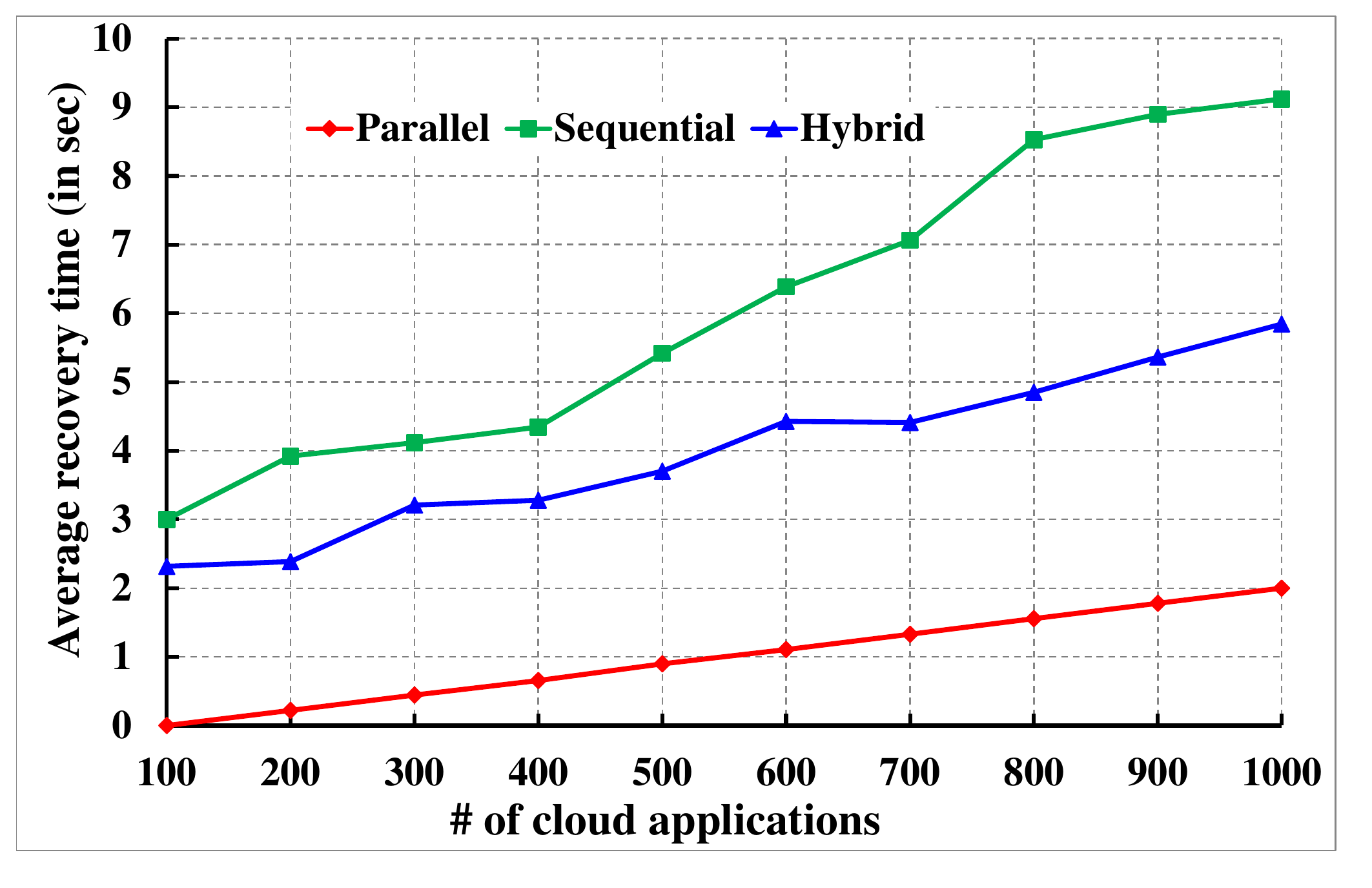}
    \vspace{-3mm}
    \caption{Recovery time under different order of execution of components.}
    \label{fig:sim:recoveryTime_ExecutionOrder}
\end{figure}

Under the proposed scheme, the components are scheduled to run
under both parallel and sequential orders. The components that are
more likely to fail are scheduled to run in a parallel manner
along with their corresponding backup components. In other words, the higher-rank backup components are scheduled to run in
parallel manner and rest of the components are scheduled to run in
the sequential manner. The effects of order of executions of the
backup components on amount of resource requirement and the
recovery time are shown in Figure
\ref{fig:sim:noofVMreq_ExecutionOrder} and
\ref{fig:sim:recoveryTime_ExecutionOrder}. Figure
\ref{fig:sim:noofVMreq_ExecutionOrder} shows the effect of the 
execution order of backup components onto the number of required VMs.
In the simulation, $5$ backup components are scheduled for each
component in case of parallel and sequential order of execution.
It is found that for $100$ cloud applications, each consisting of
$4$ to $16$ number of components, a minimum of $24$,
maximum of $96$, and an average of $51.11$ numbers of VMs required per application, when
the backup components are scheduled to run in concurrent manner.
However, the number of VMs falls to an average of $10.22$
numbers of VMs, when all the backup components are scheduled to
run in sequential manner. The minimum and maximum numbers of VMs
required in sequential manner are $4$ and $16$, respectively,
which is equal to the minimum and maximum number of components
present in a cloud application. In hybrid mode, the additional
requirement of resources in terms of number of VMs is higher than
that of the sequential order but lesser than that of the parallel
order as backup components are scheduled to run in both parallel
and sequential mode. The minimum, maximum, and the average number
of VMs required for $100$ numbers of cloud applications are $9,
96,$ and $32.57$. In the simulation setup, the backup components
of at least one component are scheduled to run in concurrent manner,
and hence a minimum of $9$ numbers of VMs is required for the
proposed hybrid order of VM execution.

The order of execution also has an impact on recovery time. The
recovery time is defined as the time required to recover the
service from the corresponding component failure. In case of the
parallel execution of the backup components, the recovery time is
very less, which ranges from $0$ to $2$ seconds for the number of
components ranging from $100$ to $1000$, as shown in Figure
\ref{fig:sim:recoveryTime_ExecutionOrder}. However, in case of
sequential execution of backup components, the recovery time is
very high as one component needs to restart from the very
beginning of its failure. The recovery time for sequential
execution of the backup components ranges from $3sec$ to $9.12sec$
for $100$ components to $1000$ components. According to the
proposed algorithm, the order of execution of backup components is
hybrid. Hence, it is observed that the recovery time for same
number of components ranges from approximately $2.3sec$ to
$5.84sec$.

\begin{figure}[h]
    \centering
    \includegraphics[width=75mm]{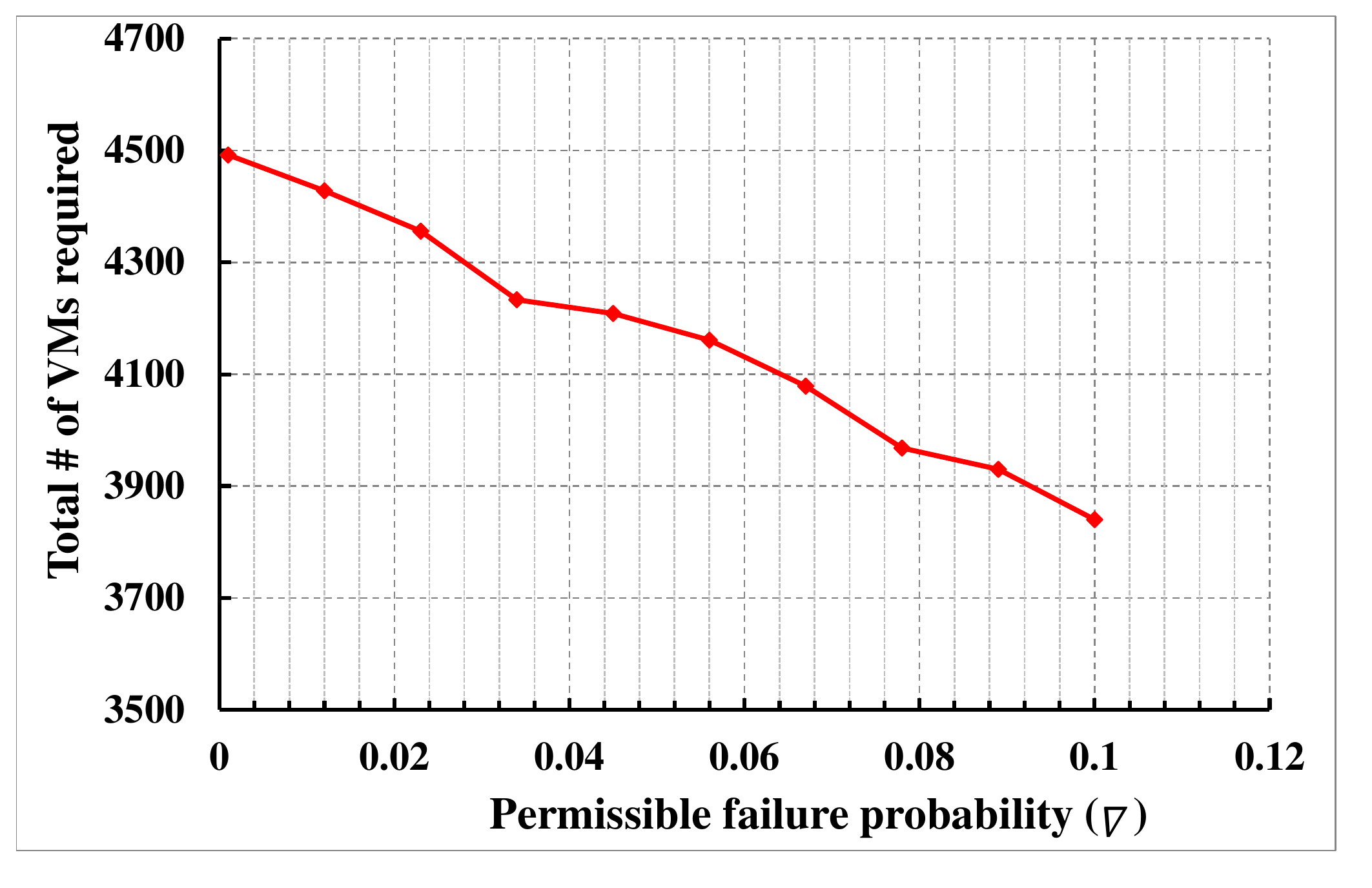}
    \vspace{-3mm}
    \caption{The impact of permissible failure probability on resource requirement.}
    \label{fig:sim:noofVMs_permisibleFP}
\end{figure}

\begin{figure}[h]
    \centering
    \includegraphics[width=75mm]{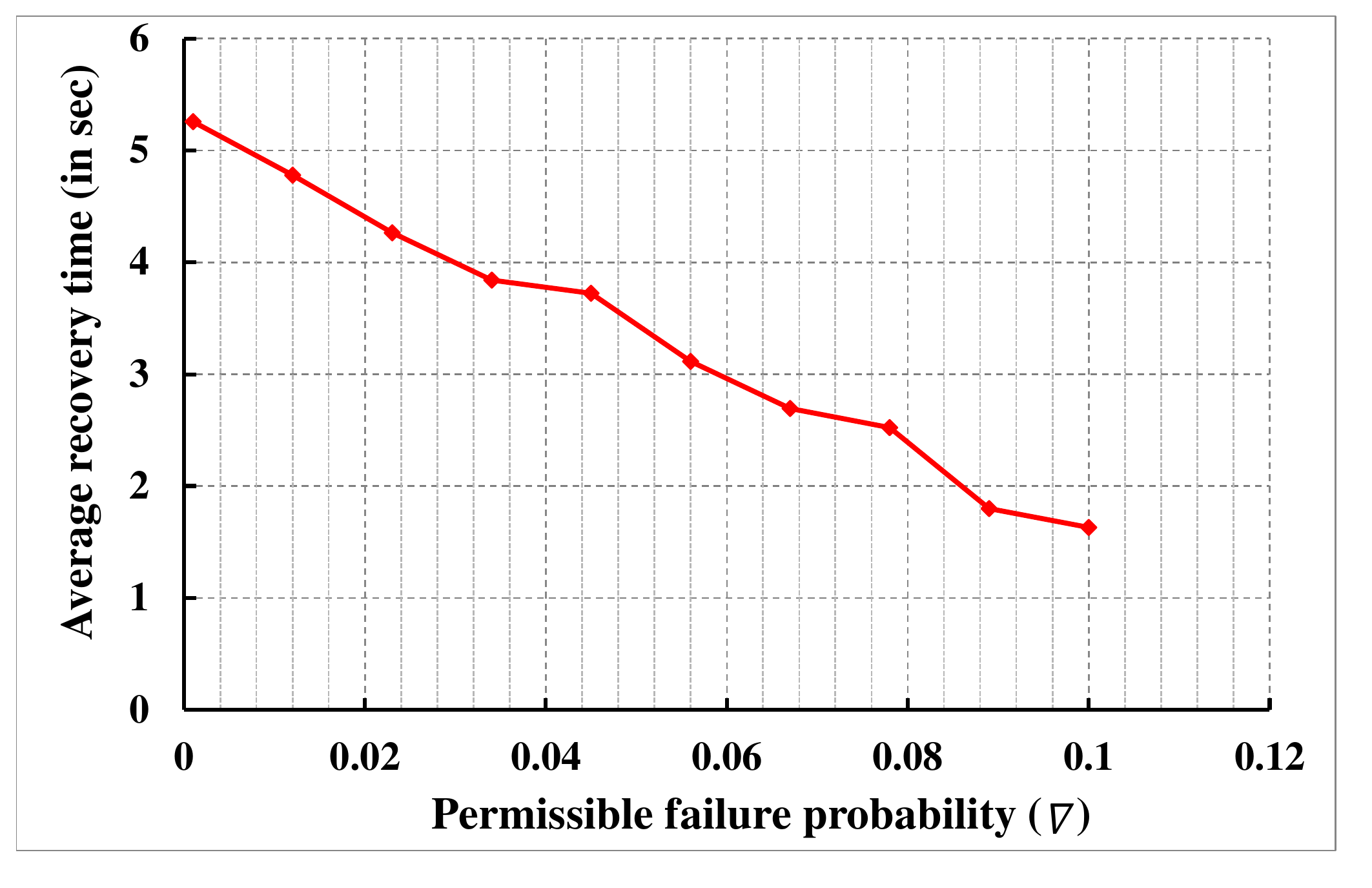}
    \vspace{-3mm}
    \caption{The impact of permissible failure probability on recovery time.}
    \label{fig:sim:recoveryTime_permisibleFP}
\end{figure}

The impacts of permissible failure probability $(\bigtriangledown)$
on the additional resource requirement and recovery time are
investigated and the results are presented in Figure
\ref{fig:sim:noofVMs_permisibleFP} and
\ref{fig:sim:recoveryTime_permisibleFP}. The backup components are
created based on the value of $(\bigtriangledown)$. Backup
components are not created for components with the failure
probability less than $(\bigtriangledown)$. Hence, the component
with higher value of $(\bigtriangledown)$ requires more number of
backup components. This is also reflected in the simulation result
presented in Figure \ref{fig:sim:noofVMs_permisibleFP}.

The amount of required resources decreases when the value of
$(\bigtriangledown)$ increases as discussed in Figure
\ref{fig:sim:noofVMs_permisibleFP}. Hence, it is obvious that the
recovery time would decrease with the increase in permissible
failure probability $(\bigtriangledown)$. The value of
$(\bigtriangledown)$ ranges from $0.001$ to $0.1$. It is observed
that the recovery time also decreases from approximately $5.25 sec$
to $1.6 sec$, when the value of $(\bigtriangledown)$ increases.
The results in Figure \ref{fig:sim:noofVMs_permisibleFP} and
\ref{fig:sim:recoveryTime_permisibleFP} are obtained by keeping
the number of cloud components to $500$.

\begin{figure}[h]
    \centering
    \includegraphics[width=75mm]{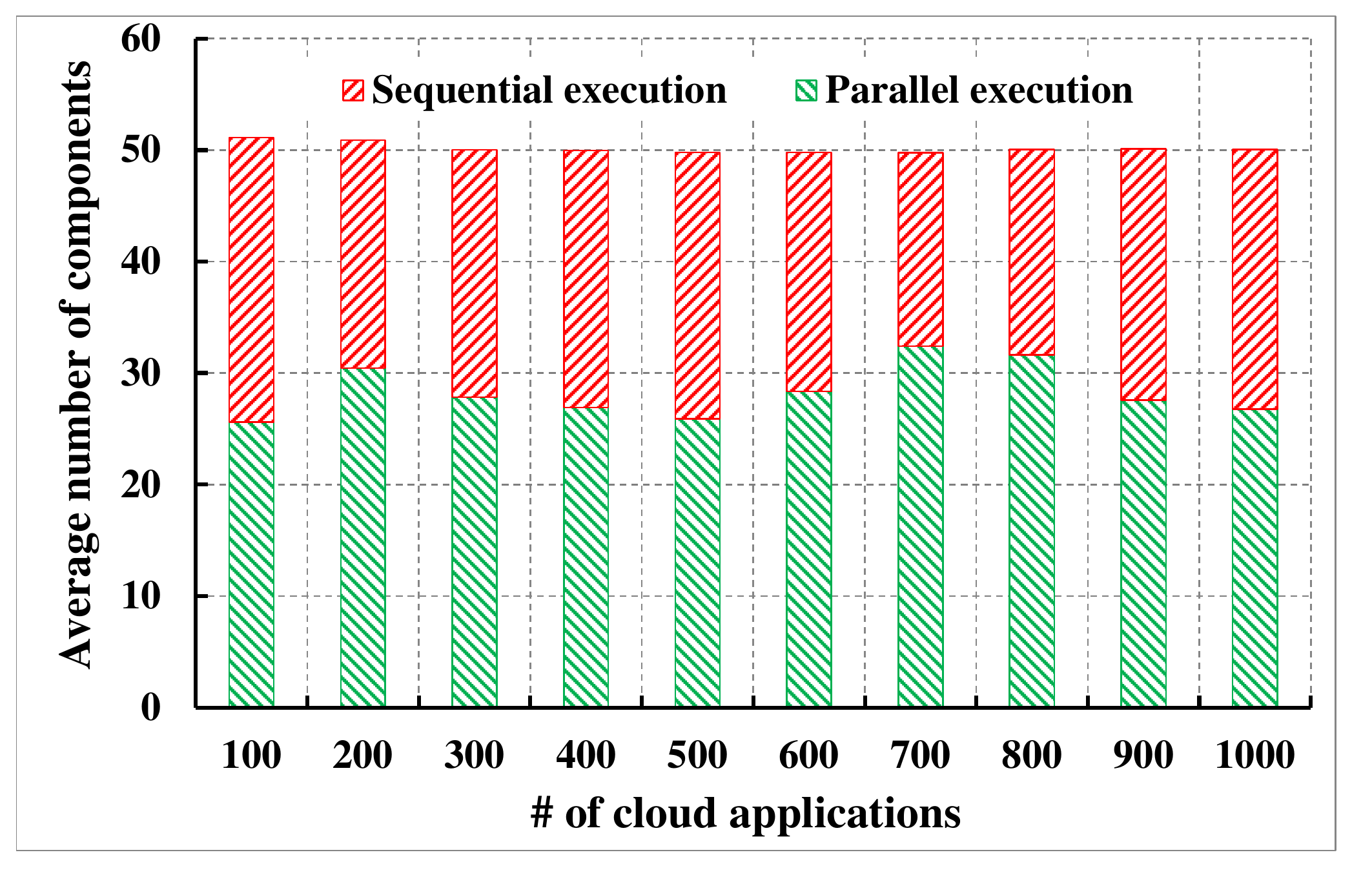}
    \vspace{-3mm}
    \caption{Number of parallel and sequential backup components.}
    \label{fig:sim:noof_backup_comps}
\end{figure}

In the proposed \emph{Hybrid k*} fault tolerant strategy, the
backup components are scheduled to run in both parallel and
sequential order. Further, the number of backup components also
varies based on the value of $\bigtriangledown$. In Figure
\ref{fig:sim:noof_backup_comps}, average number of components
along Y-axis refers to as the sum of number of primary and backup
components. For $100$ number of cloud applications, the average
number of backup components scheduled to run in parallel and
sequential order is $25.59$ and $25.51$, respectively. However,
when the number of cloud application increases to $1000$, the
average number of backup components running in parallel and
sequential order is $26.76$ and $23.30$, respectively. It is
observed that the number of backup components running in parallel
and sequential order are almost similar. Here, the number of
primary components for each cloud application ranges from $4$ to
$15$, with an average of $10$.

\begin{figure}[h]
    \centering
    \includegraphics[width=75mm]{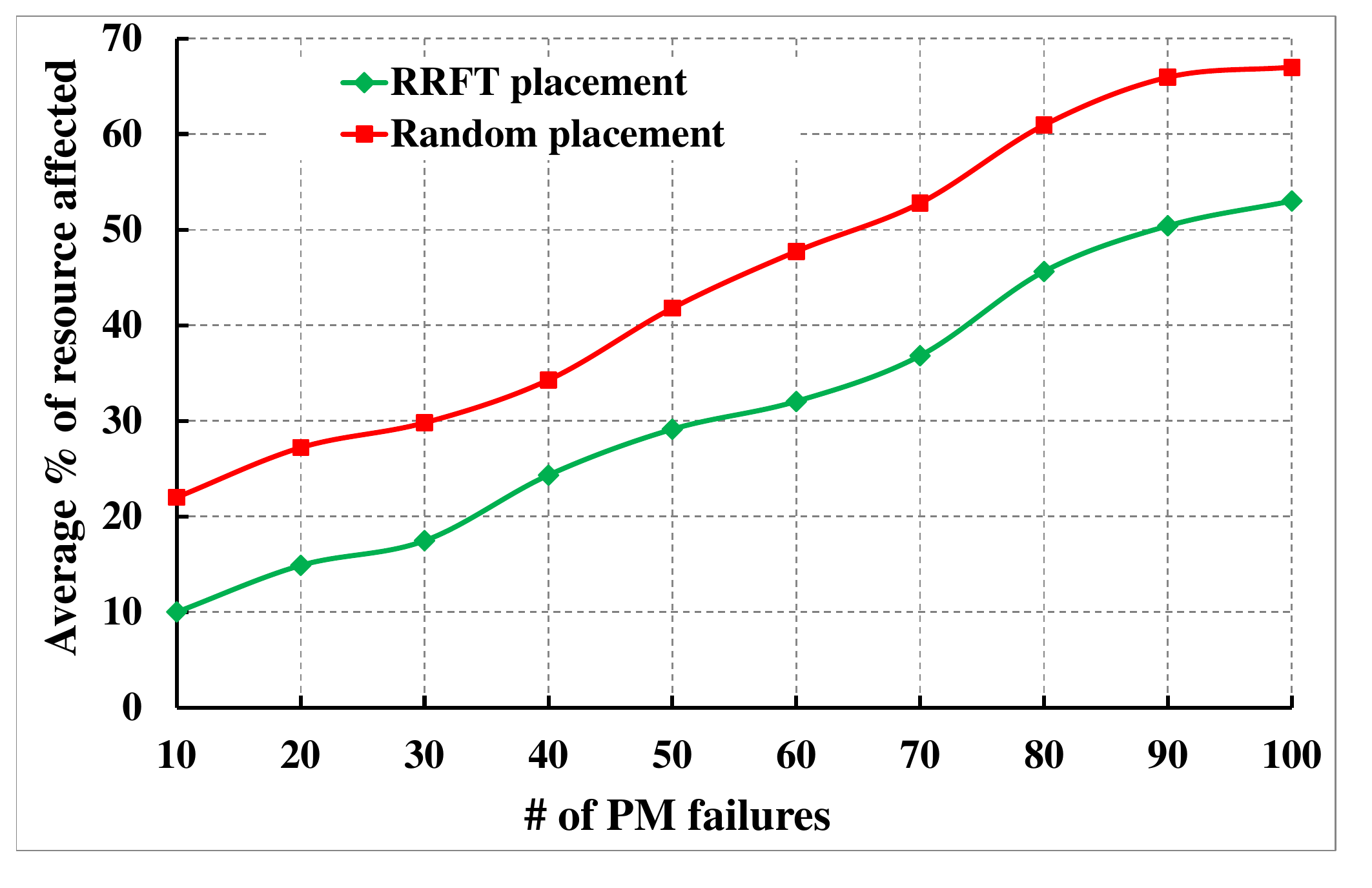}
    \vspace{-3mm}
    \caption{The impact of PM failures onto the allocated virtual resources.}
    \label{fig:sim:placemnt_resrcAffected}
\end{figure}

Considering the placement strategy as one of the performance
metrics, the effect of random placement and the proposed component
placement scheme onto the impact of PM failures are compared and
presented in Figure \ref{fig:sim:placemnt_resrcAffected}. A single
PM failure may bring down multiple primary and backup components.
The number of PM failures ranges from $10$ to $100$. The average
percentage of resource affected is calculated by obtaining the
ratio of number of VMs failed and the total number of VM currently
allocated to each cloud application. During the simulation, the
number of cloud applications is set to be $1000$. When the
components are placed onto random physical machines, it is
observed that the percentage of the resource affected is very high
as compared to that of the proposed component placement scheme.
The percentage of resource affected ranges from $22\%$ to $67\%$,
when the components are placed randomly. However, the percentage
of resource failure ranges between $10\%$ to $53\%$ when the
components are placed by following the proposed component
placement scheme, which is at least 20\% lesser over the random placement of the component.

\begin{figure}[h]
    \centering
    \includegraphics[width=75mm]{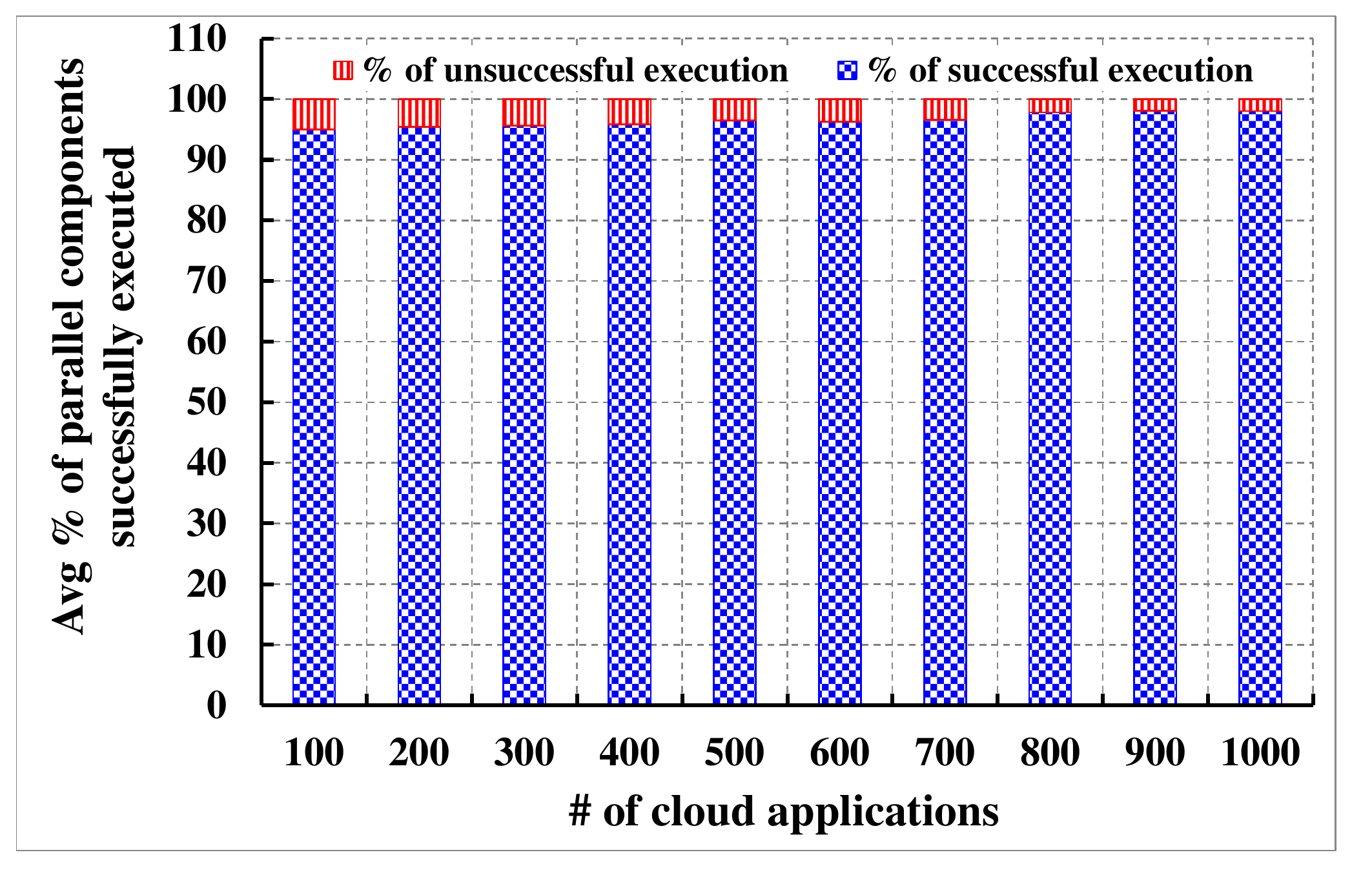}
    \vspace{-3mm}
    \caption{Percentage of parallel backup components successfully executed.}
    \label{fig:sim:successfulExecution_parallel}
\end{figure}

As a subset of the backup components is scheduled to run in
parallel order, it is essential to examine if those backup
components are executed successfully. The percentage of backup
components that are running concurrently and have finished their
execution successfully is presented in Figure
\ref{fig:sim:successfulExecution_parallel}. In other words, the
values in Y-axis represents the percentage of concurrently
scheduled backup components that are executed successfully and
unsuccessfully. It is
observed from the simulation results that the percentage of backup
components executed successfully ranges between $95\%$ to $98\%$,
when the number of cloud components ranges between $100$ and
$1000$. Hence, the percentage of components that are unable to
finish the execution ranges from $5\%$ to $2\%$.

\begin{figure}[h]
    \centering
    \includegraphics[width=75mm]{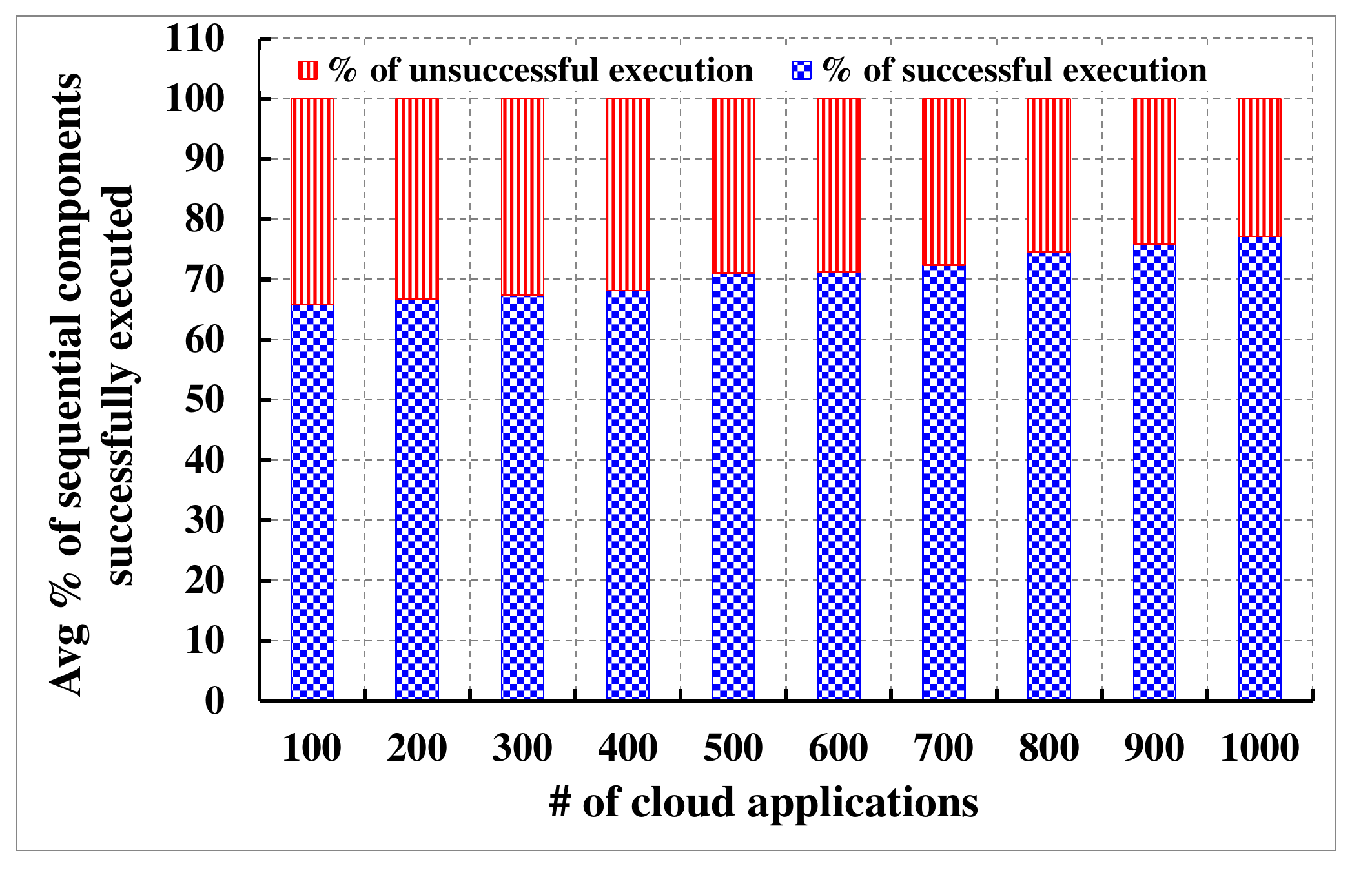}
    \vspace{-3mm}
    \caption{Percentage of sequential backup components successfully executed.}
    \label{fig:sim:successfulExecution_sequential}
\end{figure}

Similarly, the percentage of backup components that are scheduled
to run in sequential order and are executed successfully is
examined and presented in Figure
\ref{fig:sim:successfulExecution_sequential}. It is observed that
approximately $66\%$ of the backup components are successfully
executed when the number of cloud components is $100$. The
percentage increases to approximately $78\%$, when the number of
cloud components increases to $1000$. We can conclude here that
with an increase in the number of cloud components, the percentage of
failure execution of sequentially scheduled backup components
decreases. However, the percentage value along Y-axis will remain
saturated when the number of cloud application increases further.

\section{Conclusions and Future works}\label{sec:conclusion}
In this paper, we addressed an important  problem of providing fault tolerant services to the cloud components based on their ranking.
As providing fault tolerant services could be resource-intensive,
our objective was to minimize the amount of additional resources
required to provide the fault tolerant services. It is assumed
that the cloud applications comprise multiple cloud components.
The significance or importance level of all cloud components with
respect to a particular cloud application is different. Hence, we
first proposed a mechanism to find out the most significant
component, based on which ranking of the component is done.
Following the ranking mechanism, \emph{Hybrid k*} fault tolerant
strategy is proposed with the objective to provide the maximum
degree of fault tolerant to the components with higher rank values
and to provide minimum amount of additional resources to the
components with lower rank values. Besides, a component placement
strategy is proposed with the objective to find a suitable PM for
each primary and backup component. The objective of proposing the
component placement strategy is also to reduce the impact of
single PM failure onto single cloud application. 

The proposed scheme is evaluated with the similar component ranking and fault
tolerance strategies. 
It is observed that the proposed RRFT algorithm reduced the required number of VMs and PMs by approximately 10\% and 4.2\%, respectively. With the proposed placement strategy, the percentage of virtual resource affected due to failure of a PM is found to be at least 20\% lesser than that of random placement strategy.
However, extending the current version of the algorithm by considering other parameters such as makespan, user priority, cost of each components etc. and experimenting the extended version of the proposed algorithm in real environment would be the part of our future works as it is hard to build a simulated environment with exact configuration of real environment. In addition to this, we will also divert our focus to investigate the difference in characteristics of VMs and containers and make the proposed algorithm more applicable to the micro-service and container-based cloud application. 

\section*{Acknowledgment}
This work is supported by the Ministry of Science and Technology (MOST), Taiwan under grant number 110-2221-E-182-008-MY3.

\bibliographystyle{ieeetran}
\bibliography{referenceFile}

\begin{thebibliography}{10}
\providecommand{\url}[1]{#1}
\csname url@samestyle\endcsname
\providecommand{\newblock}{\relax}
\providecommand{\bibinfo}[2]{#2}
\providecommand{\BIBentrySTDinterwordspacing}{\spaceskip=0pt\relax}
\providecommand{\BIBentryALTinterwordstretchfactor}{4}
\providecommand{\BIBentryALTinterwordspacing}{\spaceskip=\fontdimen2\font plus
\BIBentryALTinterwordstretchfactor\fontdimen3\font minus
  \fontdimen4\font\relax}
\providecommand{\BIBforeignlanguage}[2]{{%
\expandafter\ifx\csname l@#1\endcsname\relax
\typeout{** WARNING: IEEEtran.bst: No hyphenation pattern has been}%
\typeout{** loaded for the language `#1'. Using the pattern for}%
\typeout{** the default language instead.}%
\else
\language=\csname l@#1\endcsname
\fi
#2}}
\providecommand{\BIBdecl}{\relax}
\BIBdecl

\bibitem{Ghazouani201761}
S.~Ghazouani and Y.~Slimani, ``A survey on cloud service description,''
  \emph{Journal of Network and Computer Applications}, vol.~91, pp. 61 -- 74,
  2017.

\bibitem{dehurydyvine}
C.~K. {Dehury} and P.~K. {Sahoo}, ``{DYVINE: Fitness-Based Dynamic Virtual
  Network Embedding in Cloud Computing},'' \emph{IEEE Journal on Selected Areas
  in Communications}, vol.~37, no.~5, pp. 1029--1045, May 2019.

\bibitem{Ding201747}
Y.~Ding, G.~Yao, and K.~Hao, ``Fault-tolerant elastic scheduling algorithm for
  workflow in cloud systems,'' \emph{Information Sciences}, vol. 393, pp. 47 --
  65, 2017.

\bibitem{Firmament2016}
I.~Gog, M.~Schwarzkopf, A.~Gleave, R.~N.~M. Watson, and S.~Hand, ``Firmament:
  Fast, centralized cluster scheduling at scale,'' in \emph{Proceedings of the
  12th USENIX Conference on Operating Systems Design and Implementation}, ser.
  OSDI'16.\hskip 1em plus 0.5em minus 0.4em\relax USA: USENIX Association,
  2016, p. 99–115.

\bibitem{FTworkflow_2017}
G.~Yao, Y.~Ding, and K.~Hao, ``Using imbalance characteristic for
  fault-tolerant workflow scheduling in cloud systems,'' \emph{IEEE
  Transactions on Parallel and Distributed Systems}, vol.~PP, no.~99, pp. 1--1,
  2017.

\bibitem{cheraghlou2016survey}
M.~N. Cheraghlou, A.~Khadem-Zadeh, and M.~Haghparast, ``A survey of fault
  tolerance architecture in cloud computing,'' \emph{Journal of Network and
  Computer Applications}, vol.~61, pp. 81--92, 2016.

\bibitem{7582315}
P.~He, X.~Zhao, C.~Tan, Z.~Zheng, and Y.~Yuan, ``Evaluation and optimization of
  the mixed redundancy strategy in cloud-based systems,'' \emph{China
  Communications}, vol.~13, no.~9, pp. 237--248, Sept 2016.

\bibitem{7742909}
M.~Amoon, ``Adaptive framework for reliable cloud computing environment,''
  \emph{IEEE Access}, vol.~4, pp. 9469--9478, 2016.

\bibitem{7096978}
W.~Chen, R.~F. da~Silva, E.~Deelman, and T.~Fahringer, ``Dynamic and
  fault-tolerant clustering for scientific workflows,'' \emph{IEEE Transactions
  on Cloud Computing}, vol.~4, no.~1, pp. 49--62, Jan 2016.

\bibitem{zheng2012component}
Z.~Zheng, T.~C. Zhou, M.~R. Lyu, and I.~King, ``Component ranking for
  fault-tolerant cloud applications,'' \emph{IEEE Transactions on Services
  Computing}, vol.~5, no.~4, pp. 540--550, 2012.

\bibitem{KUMARI2018}
P.~Kumari and P.~Kaur, ``A survey of fault tolerance in cloud computing,''
  \emph{Journal of King Saud University - Computer and Information Sciences},
  2018.

\bibitem{HASAN2018156}
M.~Hasan and M.~S. Goraya, ``Fault tolerance in cloud computing environment: A
  systematic survey,'' \emph{Computers in Industry}, vol.~99, pp. 156--172,
  2018.

\bibitem{dehurylvrm}
P.~K. {Sahoo}, C.~K. {Dehury}, and B.~{Veeravalli}, ``Lvrm: On the design of
  efficient link based virtual resource management algorithm for cloud
  platforms,'' \emph{IEEE Transactions on Parallel and Distributed Systems},
  vol.~29, no.~4, pp. 887--900, April 2018.

\bibitem{8036409}
G.~{Fan}, L.~{Chen}, H.~{Yu}, and D.~{Liu}, ``Modeling and analyzing dynamic
  fault-tolerant strategy for deadline constrained task scheduling in cloud
  computing,'' \emph{IEEE Transactions on Systems, Man, and Cybernetics:
  Systems}, vol.~50, no.~4, pp. 1260--1274, April 2020.

\bibitem{2018-fault-scheduling}
H.~Han, W.~Bao, X.~Zhu, X.~Feng, and W.~Zhou, ``Fault-tolerant scheduling for
  hybrid real-time tasks based on cpb model in cloud,'' \emph{IEEE Access},
  vol.~6, pp. 18\,616--18\,629, 2018.

\bibitem{SETLUR202014}
A.~R. Setlur, S.~J. Nirmala, H.~S. Singh, and S.~Khoriya, ``An efficient fault
  tolerant workflow scheduling approach using replication heuristics and
  checkpointing in the cloud,'' \emph{Journal of Parallel and Distributed
  Computing}, vol. 136, pp. 14 -- 28, 2020.

\bibitem{yao2017using}
G.~Yao, Y.~Ding, and K.~Hao, ``Using imbalance characteristic for
  fault-tolerant workflow scheduling in cloud systems,'' \emph{IEEE
  Transactions on Parallel and Distributed Systems}, vol.~28, no.~12, pp.
  3671--3683, 2017.

\bibitem{2018-fault-TPDS}
Z.~Wang, L.~Gao, Y.~Gu, Y.~Bao, and G.~Yu, ``A fault-tolerant framework for
  asynchronous iterative computations in cloud environments,'' \emph{IEEE
  Transactions on Parallel and Distributed Systems}, vol.~29, no.~8, pp.
  1678--1692, Aug 2018.

\bibitem{dean2016perfcompass}
D.~J. Dean, H.~Nguyen, P.~Wang, X.~Gu, A.~Sailer, and A.~Kochut, ``Perfcompass:
  Online performance anomaly fault localization and inference in
  infrastructure-as-a-service clouds,'' \emph{IEEE Transactions on Parallel and
  Distributed Systems}, vol.~27, no.~6, pp. 1742--1755, 2016.

\bibitem{8964469}
B.~{Ray}, A.~{Saha}, S.~{Khatua}, and S.~{Roy}, ``Proactive fault-tolerance
  technique to enhance reliability of cloud service in cloud federation
  environment,'' \emph{IEEE Transactions on Cloud Computing}, pp. 1--1, 2020.

\bibitem{YUAN2020106953}
G.~Yuan, Z.~Xu, B.~Yang, W.~Liang, W.~K. Chai, D.~Tuncer, A.~Galis, G.~Pavlou,
  and G.~Wu, ``Fault tolerant placement of stateful vnfs and dynamic fault
  recovery in cloud networks,'' \emph{Computer Networks}, vol. 166, p. 106953,
  2020.

\bibitem{deng2015computation}
S.~Deng, L.~Huang, J.~Taheri, and A.~Y. Zomaya, ``Computation offloading for
  service workflow in mobile cloud computing,'' \emph{IEEE Transactions on
  Parallel and Distributed Systems}, vol.~26, no.~12, pp. 3317--3329, 2015.

\bibitem{qiu2014reliability}
W.~Qiu, Z.~Zheng, X.~Wang, X.~Yang, and M.~R. Lyu, ``Reliability-based design
  optimization for cloud migration,'' \emph{IEEE Transactions on Services
  Computing}, vol.~7, no.~2, pp. 223--236, 2014.

\bibitem{2018-cost-fault}
Y.~Wang, Q.~He, D.~Ye, and Y.~Yang, ``Formulating criticality-based
  cost-effective fault tolerance strategies for multi-tenant service-based
  systems,'' \emph{IEEE Transactions on Software Engineering}, vol.~44, no.~3,
  pp. 291--307, March 2018.

\bibitem{smara2016acceptance}
M.~Smara, M.~Aliouat, A.-S.~K. Pathan, and Z.~Aliouat, ``Acceptance test for
  fault detection in component-based cloud computing and systems,''
  \emph{Future Generation Computer Systems}, 2016.

\bibitem{wang2016fd4c}
T.~Wang, W.~Zhang, C.~Ye, J.~Wei, H.~Zhong, and T.~Huang, ``Fd4c: Automatic
  fault diagnosis framework for web applications in cloud computing,''
  \emph{IEEE Transactions on Systems, Man, and Cybernetics: Systems}, vol.~46,
  no.~1, pp. 61--75, 2016.

\bibitem{chen2015energy}
C.-A. Chen, M.~Won, R.~Stoleru, and G.~G. Xie, ``Energy-efficient
  fault-tolerant data storage and processing in mobile cloud,'' \emph{IEEE
  Transactions on cloud computing}, vol.~3, no.~1, pp. 28--41, 2015.

\bibitem{gupta2016power}
P.~Gupta and S.~Ghrera, ``Power and fault aware reliable resource allocation
  for cloud infrastructure,'' \emph{Procedia Computer Science}, vol.~78, pp.
  457--463, 2016.

\bibitem{wang2015festal}
J.~Wang, W.~Bao, X.~Zhu, L.~T. Yang, and Y.~Xiang, ``Festal: fault-tolerant
  elastic scheduling algorithm for real-time tasks in virtualized clouds,''
  \emph{IEEE Transactions on Computers}, vol.~64, no.~9, pp. 2545--2558, 2015.

\bibitem{7976777}
K.~Vinay and S.~M.~D. Kumar, ``Fault-tolerant scheduling for scientific
  workflows in cloud environments,'' in \emph{2017 IEEE 7th International
  Advance Computing Conference (IACC)}, Jan 2017, pp. 150--155.

\bibitem{anderson1976reliability}
R.~T. Anderson, ``Reliability design handbook,'' RELIABILITY ANALYSIS CENTER
  GRIFFISS AFB NY, Tech. Rep., 1976.

\bibitem{9112332}
M.~Shifrin, R.~Mitrany, E.~Biton, and O.~Gurewitz, ``Vm scaling and load
  balancing via cost optimal mdp solution,'' \emph{IEEE Transactions on Cloud
  Computing}, pp. 1--1, 2020.

\bibitem{LIANG2021101991}
H.~Liang, X.~Wen, Y.~Liu, H.~Zhang, L.~Zhang, and L.~Wang, ``Logistics-involved
  qos-aware service composition in cloud manufacturing with deep reinforcement
  learning,'' \emph{Robotics and Computer-Integrated Manufacturing}, vol.~67,
  p. 101991, 2021.

\bibitem{li2019energy}
C.~Li, Y.~Wang, Y.~Chen, and Y.~Luo, ``Energy-efficient fault-tolerant replica
  management policy with deadline and budget constraints in edge-cloud
  environment,'' \emph{Journal of Network and Computer Applications}, vol. 143,
  pp. 152--166, 2019.

\end{thebibliography}

\begin{IEEEbiography}[{\includegraphics[width=1in,height=1.25in,clip,keepaspectratio]{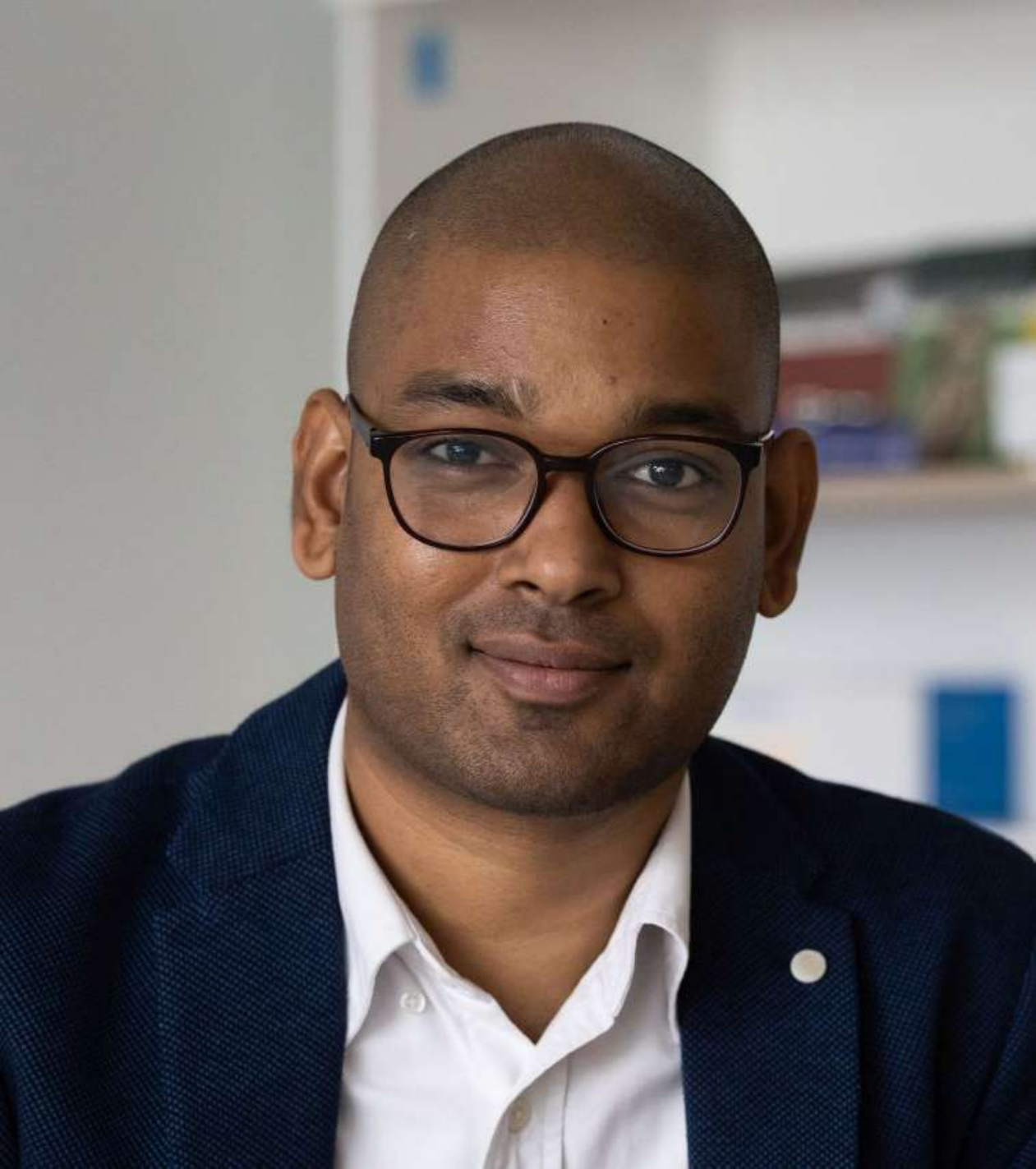}}]{Chinmaya Kumar Dehury}
Chinmaya Kumar Dehury received bachelor degree from Sambalpur University, India, in June 2009 and MCA degree from Biju Pattnaik University of Technology, India, in June 2013. He received the PhD Degree in the department of Computer Science and Information Engineering, Chang Gung University, Taiwan. He is currently a Lecturer of Distributed System, member of Mobile \& Cloud Lab in the Institute of Computer Science, University of Tartu, Estonia. His research interests include scheduling, resource management and fault tolerance problems of Cloud and fog Computing, and the application of artificial intelligence in cloud management. He is a member, IEEE, and also reviewer to several journals and conferences, such as IEEE TPDS, IEEE JSAC, Wiley Software: Practice and Experience, etc. 
\end{IEEEbiography}
\vspace{-15mm}

\begin{IEEEbiography}[{\includegraphics[width=1in,height=1.25in,clip,keepaspectratio]{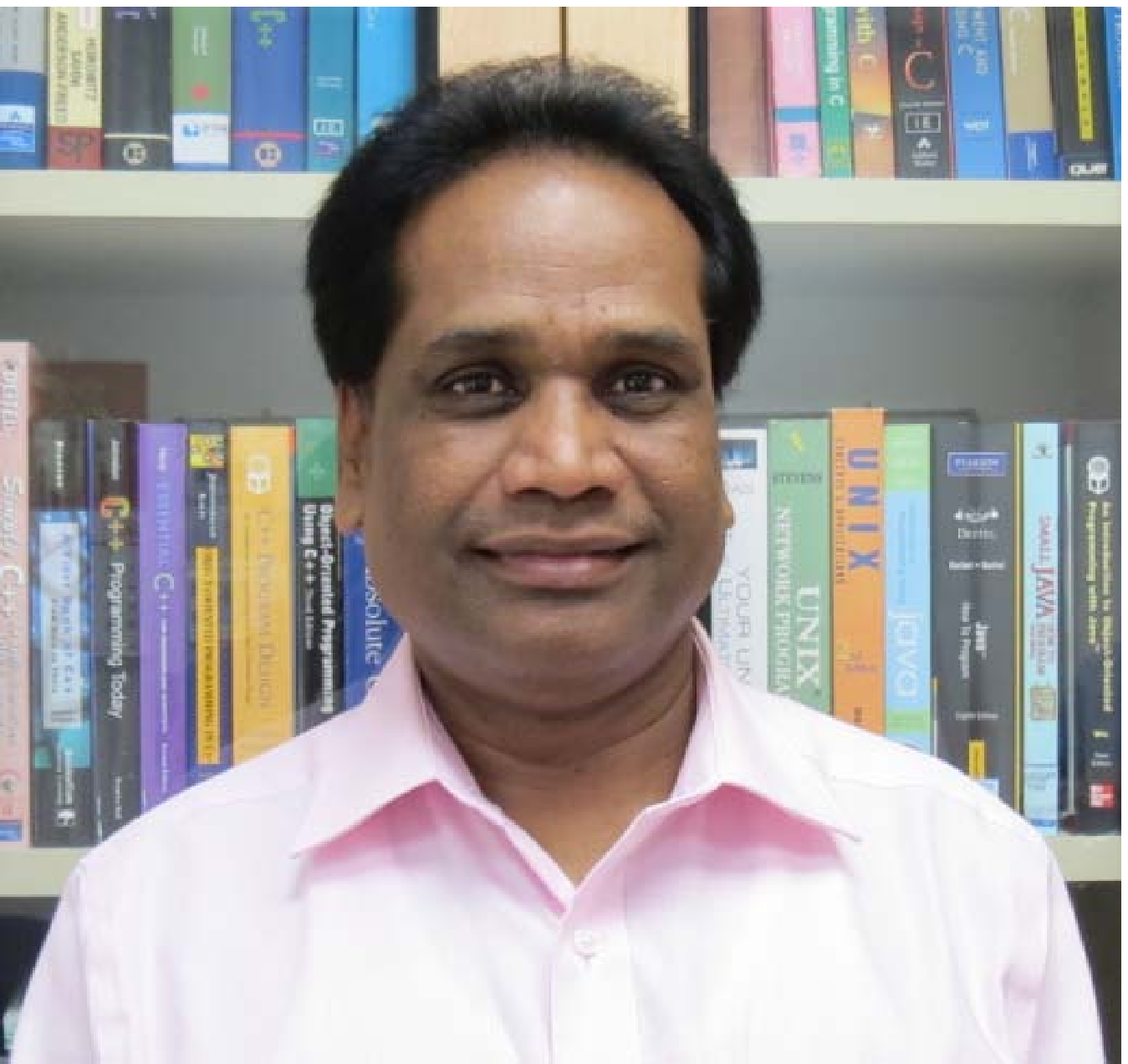}}]{Prasan Kumar Sahoo}
Prasan Kumar Sahoo received the B.Sc. degree in physics (with
Honors), the M.Sc. degree in mathematics both from Utkal
University, Bhubaneswar, India, in 1987 and 1994, respectively. He
received the M.Tech. degree in computer science from the Indian
Institute of Technology (IIT), Kharagpur, India, in 2000, the
first Ph.D. degree in mathematics from Utkal University, in 2002,
and the second Ph.D. degree in computer science and information
engineering from the National Central University, Taiwan, in 2009.
He is currently a Professor in the Department of Computer Science
and Information Engineering, Chang Gung University, Taiwan. He was an Adjunct Associate Researcher in the Department of
Cardiology and was an Adjunct Researcher in the Division of Colon and Rectum Cancer, Chang Gung Memorial Hospital, Linkou, Taiwan. Currently, he is an Adjunct Researcher in the Department of Neurology, Chang Gung Memorial Hospital, Linkou, Taiwan . He has worked as an Associate Professor in the Department of Information Management, Vanung University,
Taiwan, from 2007 to 2011. He was a Visiting Associate Professor in the Department of Computer Science, Universite Claude Bernard Lyon 1, Villeurbanne,
France. His current research interests include artificial
intelligence, big data analytic, cloud computing, and IoT. He is an Editorial Board Member of Journal of Network and Computer Applications, International Journal of Vehicle
Information and Communication Systems and was Lead Guest Editor, special issue of Electronics journal. He has worked as the
Program Committee Member of several IEEE and ACM conferences and is a senior member, IEEE.
\end{IEEEbiography}

\vspace{-15mm}
\begin{IEEEbiography}[{\includegraphics[width=1in,height=1.25in,clip,keepaspectratio]{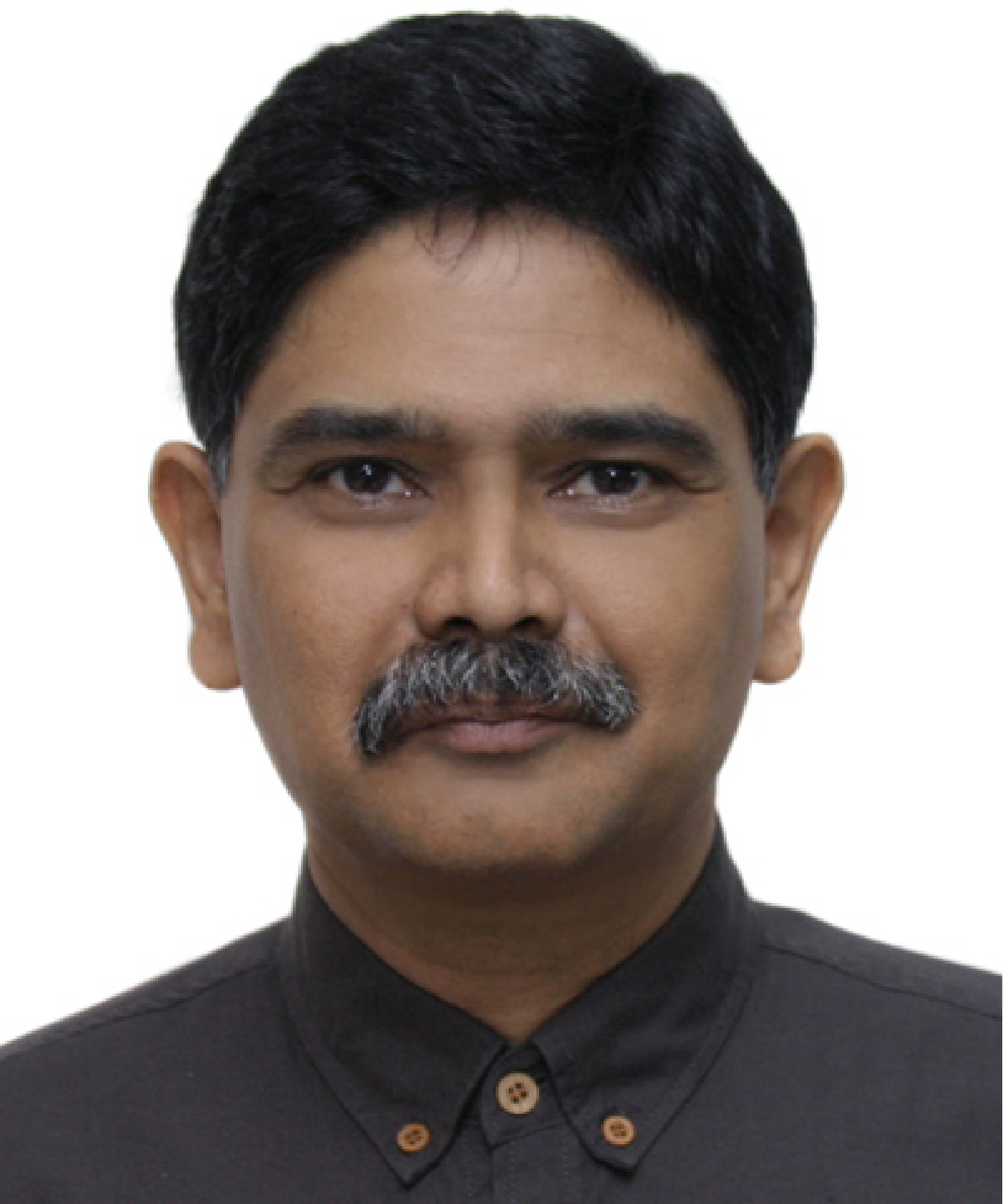}}]{Bharadwaj Veeravalli}
Bharadwaj Veeravalli received his BSc degree in Physics, from
Madurai-Kamaraj University, India, in 1987, the Master's degree in
Electrical Communication Engineering from the Indian Institute of
Science, Bangalore, India in 1991, and the PhD degree from the
Department of Aerospace Engineering, Indian Institute of Science,
Bangalore, India, in 1994. He received gold medals for his
bachelor degree overall performance and for an outstanding PhD
thesis (IISc, Bangalore India) in the years 1987 and 1994,
respectively. He is currently with the Department of Electrical
and Computer Engineering, Communications and Information
Engineering (CIE) division, at The National University of
Singapore, Singapore, as a tenured Associate Professor. His main
stream research interests include cloud/grid/cluster computing
(big data processing, analytics and resource allocation),
scheduling in parallel and distributed systems, Cybersecurity, and
multimedia computing. He is one of the earliest researchers in the
field of Divisible Load Theory (DLT). He is currently serving the
editorial board of IEEE Transactions on Parallel and Distributed
Systems as an associate editor. He is a senior member of the IEEE
and the IEEE-CS.
\end{IEEEbiography}

\end{document}